\documentclass[journal]{IEEEtran} 
\usepackage{amsmath,amsfonts}
\usepackage{algorithmic}
\usepackage{algorithm}
\usepackage{array}
\usepackage{textcomp}
\usepackage{stfloats}
\usepackage{url}
\usepackage{verbatim}
\usepackage{graphicx}
\usepackage{cite}

\usepackage{adjustbox}

\usepackage{subfigure}

\usepackage{array}

\usepackage{booktabs} 
\usepackage{amsmath}  

\usepackage{calc} 
\usepackage{cuted} 
\usepackage{mathtools} 

\usepackage{diagbox} 
\usepackage{tabularx}
\newcolumntype{Y}{>{\centering\arraybackslash}X} 

\usepackage{xcolor}
\usepackage{graphicx}
\usepackage[colorlinks=true, linkcolor=blue, urlcolor=blue, citecolor=blue]{hyperref} 

\newcommand{\bluecite}[1]{\textcolor{blue}{\cite{#1}}}

\makeatletter
\newcommand{\citenew}[1]{{\def\@cite##1##2{##1\if@tempswa , ##2\fi}\cite{#1}}}
\makeatother

\usepackage{booktabs}              
\usepackage{makecell}              
\usepackage{amssymb}

\begin{document}

\title{Vertical Pinching Antenna Systems (V-PAS) \\Aided UAV Communications}

\author{Tao Wang, 
        Kunrui Cao,
        Yayun Qu, 
        Lu Lv,~\IEEEmembership{Member,~IEEE}, 
        Xianfu Lei,~\IEEEmembership{Member,~IEEE},
        \\Dimitrios Tyrovolas,~\IEEEmembership{Member,~IEEE},
        and Panagiotis D. Diamantoulakis,~\IEEEmembership{Senior Member,~IEEE}
\thanks{Tao Wang is with the School of Information and Communications, National University of Defense Technology, Wuhan 430035, China, and also with the Information Support Force Engineering University, Wuhan 430035, China (e-mail: twang@nudt.edu.cn).}
\thanks{Kunrui Cao and Yayun Qu are with the Information Support Force Engineering University, Wuhan 430035, China, and also with the School of Information and Communications, National University of Defense Technology, Wuhan 430035, China (e-mail: krcao@nudt.edu.cn; quyayun@nudt.edu.cn).}
\thanks{Lu Lv is with the State Key Laboratory of Integrated Services Networks, Xidian University, Xi’an 710071, China (e-mail: lulv@xidian.edu.cn).}
\thanks{Xianfu Lei is with the School of Information Science and Technology, Southwest Jiaotong University, Chengdu 610031, China (e-mail: xflei@swjtu.edu.cn).}
\thanks{Dimitrios Tyrovolas is with the Department of Informatics and 
Telecommunications, University of Peloponnese, 22132, Tripoli, Greece 
(email: d.tyrovolas@uop.gr)}
\thanks{Panagiotis D. Diamantoulakis is with the Department of Electrical and Computer Engineering, Aristotle University of Thessaloniki, 54124 Thessaloniki, Greece (e-mail: padiaman@auth.gr).}
}

\markboth{}%
{Shell \MakeLowercase{\textit{et al.}}: A Sample Article Using IEEEtran.cls for IEEE Journals}

\IEEEaftertitletext{\vspace{-1.5\baselineskip}}

\maketitle
\vfill
\begin{abstract}
To address the limitation that existing horizontal pinching antenna systems (H-PAS) are primarily designed for ground users and confined to two-dimensional (2D) coverage, 
this paper proposes deploying a pinching antenna (PA) along the facade of urban buildings to construct a vertical pinching antenna system (V‑PAS) for ultra‑low‑altitude unmanned aerial vehicle (UAV) communications.
The proposed architecture employs a dielectric waveguide continuously deployed along the full height of buildings, extending the considered user service region of PAS from a 2D ground plane to a three-dimensional (3D) low-altitude airspace, enabling stable line-of-sight (LoS) links for “last meter” UAV communications.
We define the concept of pinching multiplicative path loss (PMPL) to characterize the cascaded multiplicative attenuation of the waveguide and free-space path losses. It is found that PMPL is insensitive to vertical distance, rendering V‑PAS highly adaptive to building heights.
Furthermore, accurate and asymptotic closed-form expressions for outage probability and ergodic rate under lossy waveguide conditions are derived, respectively. 
The optimal access point (AP) height and PA position are further characterized, providing deployment insights for practical V-PAS.
The results show that V‑PAS achieves performance advantages over H-PAS and the benchmark without PA in most ultra-low-altitude UAV communication scenarios.
Only under extremely high transmission power and large UAV operational area may the outage probability of V‑PAS with a lossy waveguide be inferior to that of the benchmark without PA, but the ergodic rate of V-PAS still maintains an advantage. By contrast, V‑PAS with a lossless waveguide outperforms the benchmark without PA in UAV communications, representing the theoretical performance upper bound of V‑PAS.
\end{abstract}


\section{Introduction}
\subsection{Background and Related Works}

\IEEEPARstart{W}{ith} the large-scale commercialization of the fifth-generation (5G), terrestrial cellular networks have achieved high-speed broadband coverage in major populated areas, supporting the rapid development of mobile communications and the Internet of Things (IoT)\bluecite{Liu1}. However, the coverage of 5G remains largely confined to the two-dimensional (2D) plane, falling short of the space-air-ground-sea integrated networks envisioned by the sixth-generation (6G)\bluecite{Vaezi},\bluecite{Wangchengxiang}. In this grand blueprint, the low-altitude economy with unmanned aerial vehicle (UAV) is rapidly emerging and is widely recognized as one of the most promising markets in 6G\bluecite{Gran}. 
From urban logistics delivery and infrastructure inspection to air traffic management and emergency communications, the widespread use of UAVs poses unprecedented challenges to wireless communication networks. These networks must possess three-dimensional (3D) sensing capabilities for low-altitude airspace, as well as ultra-reliable and low-latency communication (URLLC) capabilities\bluecite{Huang},\bluecite{ADD1}. 
However, widely deployed cellular networks were originally designed to serve terrestrial users, and exhibit fundamental performance limitations when supporting low-altitude UAV communications\bluecite{Jiang},\bluecite{ADD2}.
The antennas of terrestrial base stations typically employ a fixed downtilt design, which concentrates the radiated energy mainly near the ground and causes sharp signal attenuation in low-altitude areas\bluecite{Zhao1}.
These limitations are rooted in the physical layer and the underlying network architecture, and cannot be overcome simply by increasing transmission power or deploying more base stations.
Therefore, it is urgent to explore novel antenna architectures and access paradigms for low-altitude coverage.

Programmable wireless environment (PWE), as a revolutionary new paradigm for achieving full-domain coverage in 6G, fundamentally overturns the conventional design philosophy where communication systems passively adapt to the wireless propagation environment\bluecite{Lia}.
The core idea of PWE lies in transforming such passive channel adaptation into active electromagnetic regulation, thereby enabling programmable control of the wireless channel and fundamentally enhancing the performance of communication systems\bluecite{Wu1}.
Reconfigurable intelligent surface (RIS) is one of the most representative PWE technologies.
Through proper phase shift optimization, RIS can substantially enhance the quality and coverage of communication links\bluecite{ZijianZhang}.
Moreover, researchers have proposed several enhanced RISs.
Simultaneously transmitting and reflecting RIS (STAR-RIS) overcomes reflection-only limitation and achieves 360-degree full-space coverage\bluecite{Chi}.
Active RIS amplifies incident signals to compensate for path loss\bluecite{Cao1}.
Zero-energy RIS harvests radio-frequency energy from incident waves to achieve self-sustainability\bluecite{Chen1}.
However, RIS suffers from a fundamental limitation, namely the “multiplicative fading" effect in cascaded channels, which causes the performance gain to degrade rapidly with increasing propagation distance.

Movable antenna (MA) and fluid antenna (FA) represent another important category of PWE, which improve channel conditions by dynamically changing the physical positions of antennas\bluecite{Zhangrui2},\bluecite{Shao}.
MAs adjust their spatial positions through motor-driven actuation\bluecite{Zhangrui1}. FAs flexibly switch their radiation positions by controlling the flow of liquid metal\bluecite{KKW1}.
These two technologies can identify the optimal channel position within a small range, effectively mitigating the performance degradation caused by small-scale fading.
However, the position adjustment range of MAs and FAs is typically limited to the order of several wavelengths. Thus, MA and FA cannot fundamentally resolve the large-scale fading problem.
Rotatable antenna (RA) represents an emerging branch of PWE technologies, which introduces an additional spatial degree of freedom through the antenna orientation angle, enabling dynamic steering of the main lobe to concentrate energy toward target users\bluecite{Zheng1}.
Compared with MA and FA, RA features a simpler hardware structure, achieving beam pointing adjustment merely by mechanical or electronic rotation\bluecite{Zheng2}.
However, RA cannot change the physical position of the antenna, which limits its ability to mitigate path loss.

To address the common limitations of the aforementioned PWE technologies, pinching antenna (PA) emerges as a breakthrough paradigm, with the concept and prototype first proposed by NTT DOCOMO in 2022\bluecite{NTT}.
The core innovation of PA lies in transforming path loss from an uncontrollable environmental parameter into an actively configurable system variable, introducing a new degree of freedom for wireless channel control.
PA employs a low-loss dielectric waveguide as the transmission backbone and uses passive dielectric elements to form controllable radiation  points along the waveguide, allowing dynamic adjustment of their number and positions, flexible configuration of the coverage area, and the establishment of stable line-of-sight (LoS) links\bluecite{Ding},\bluecite{Yangzheng}.
Consequently, PA can effectively overcome large-scale path loss and enhance the reliability of “last meter" communication scenarios\bluecite{Cao2}.
Meanwhile, PA supports flexible pinching beamforming, enabling precise control of signal strength and interference\bluecite{D},\bluecite{Wangzhaolin}.
In addition, PA features a simple structure, low deployment cost, and high scalability\bluecite{How}.
These characteristics make PA inherently suitable for scenarios such as ultra-low-altitude UAV communications, positioning PA as one of the most promising research directions in 6G programmable wireless environments.
\subsection{Motivation and Contributions}
Recently, some initial studies have explored the potential of the pinching antenna system (PAS), but related research still remains in its infancy.
The authors of\bluecite{Xu} proposed a low-complexity algorithm for rate maximization in the downlink PAS and verified its performance advantage.
In\bluecite{WangKaidi}, the authors proposed a matching-based antenna activation scheme for non-orthogonal multiple access (NOMA) assisted PAS.
In\bluecite{Ouyang}, the authors analyzed the performance limits of PA-assisted integrated sensing and communication (ISAC) systems.a
Moreover, the authors of\bluecite{Ding2} proposed environment division multiple access (EDMA) using PA to reconfigure LoS links and suppress interference.
Recent advances have also explored multi-waveguide and multi-user PAS architectures. In\bluecite{ZhaoJJ1}, the authors proposed and compared three transmission structures for multi-waveguide PAS, namely waveguide multiplexing (WM), waveguide division (WD), and waveguide switching (WS). Building upon the WD structure, the authors of\bluecite{ZhaoJJ2} introduced waveguide division multiple access (WDMA), which assigns a dedicated waveguide to each user and exploits pinching beamforming for near orthogonality in free space. Furthermore, in\bluecite{ZhaoJJ3}, the authors proposed a WD-based NOMA framework for  multi-user scenarios and studied the joint optimization of power allocation, pinching beamforming, and user scheduling.
Recent studies have further extended PAS in different architectural dimensions.
In\bluecite{WKD1}, an LCX-based generalized PAS with dual-port feeding was proposed to enable bidirectional guided-wave excitation and provide additional spatial degrees of freedom. In\bluecite{WKD2}, adjustable slot apertures were introduced to generalize binary slot activation to continuous aperture adjustment. 
The authors of\bluecite{WKD3} considered finite PA movement speed and proposed antenna roaming, where the PA continuously moves along the waveguide while providing communication service. These studies enhance PAS from the perspectives of feeding architecture, radiation control, and antenna movement, respectively.

Despite these recent advances in PAS, existing analytical frameworks still predominantly assume horizontal waveguide deployment for ground users.
More specifically, existing horizontal pinching antenna systems (H-PAS) typically employ dielectric waveguides deployed at a fixed height, along which the PA is adjusted in the horizontal direction to match users distributed mainly over a 2D plane.
This deployment naturally fits horizontally extended service environments, such as indoor ceiling deployments and corridors, where user distribution and mobility are dominated by the horizontal dimension. In contrast, UAVs operating around high-rise buildings may move over a wide range of altitudes. For such vertically extended service regions, a single H-PAS deployed at a fixed height cannot continuously follow the UAV altitude over the full building height, while the vertical pinching antenna system (V-PAS) directly exploits a dielectric waveguide vertically mounted along the building facade, and enables the PA to be adjusted along the vertical direction to follow the UAV altitude. Therefore, V-PAS provides a new architecture whose PA adjustment direction is naturally matched to the dominant vertical mobility of UAV.
No prior work has investigated the feasibility of vertically deploying PA on building facades to construct the vertical pinching antenna system (V-PAS), let alone fully considered the unique vertical-dimension channel characteristics of ultra-low-altitude UAV communications. 
Specifically, the V-PAS introduces new challenges, including 3D channel modeling, the cascaded multiplicative attenuation of the waveguide and free-space path losses, and the optimization of the access point (AP) height, which cannot be directly addressed by existing horizontal PA architectures.
To address these critical research gaps, this paper first proposes the vertical deployment of PA on building facades to construct a V-PAS for ultra-low-altitude UAV communications, achieving efficient and reliable wireless transmissions under practical deployment conditions. The main contributions of this paper are summarized as follows.
\begin{itemize}
\item
We propose vertically deploying PA along the facade of buildings to construct the V-PAS architecture for ultra-low-altitude UAV communications.
The proposed architecture extends the considered user service region of PAS from a 2D ground plane to a 3D low-altitude airspace. By exploiting a dielectric waveguide continuously deployed along the full height of urban buildings, it ensures a stable LoS link and overcomes the limitation of the existing H-PAS that only serves ground users. To the best of the authors’ knowledge, this is the first work to explore the use of V-PAS for UAV communications.

\item
Based on the physical characteristics of the lossy dielectric waveguide, we formally define the concept of \textit{pinching multiplicative path loss (PMPL)}, which captures the cascaded multiplicative attenuation of the waveguide and free-space path losses in vertical deployment scenarios. 
We find that PMPL is insensitive to vertical distance, making V-PAS highly adaptive to building heights. The performance advantage region of V-PAS is further characterized quantitatively, providing clear guidance for its deployment in “last meter” communication scenarios.

\item
Considering that the UAV is uniformly distributed within a 3D cuboid airspace, accurate and asymptotic closed-form expressions for outage probability and ergodic rate under a lossy waveguide are derived. We further find and prove that the system performance of V-PAS exhibits midpoint symmetry with respect to AP height, and both the optimal outage probability and ergodic rate are achieved at the midpoint. 
Moreover, the optimal PA position that maximizes the received SNR is derived by balancing the waveguide loss and free-space path loss.

\item
Theoretical and numerical results reveal the impact of key parameters including transmission power, waveguide absorption coefficient, AP height and UAV operational area on system performance. They validate that V-PAS outperforms H-PAS and the benchmark without PA in most ultra-low-altitude UAV communication scenarios. Only under extremely high transmission power and large UAV operational area may the outage probability of V-PAS with a lossy waveguide be inferior to that of the benchmark without PA, but the ergodic rate of V-PAS remains superior. By contrast, V-PAS with a lossless waveguide consistently outperforms the benchmark without PA in UAV communications under all conditions.

\end{itemize}
\subsection{Organization}
The remainder of this paper is organized as follows. In Section \ref{section:2}, the V‑PAS architecture for UAV communications is introduced, followed by the channel model and signal model. In Section \ref{section:3}, we analyze the performance of UAV communications achieved by the V-PAS, and derive accurate and asymptotic closed-form expressions for outage probability and ergodic rate. 
In Section \ref{section:4}, we define the concept of PMPL and investigate the deployment optimality of AP and PA.
In Section \ref{section:5}, the simulation results and discussions are presented to gain deep insights. Section \ref{section:6} concludes the paper. 
\begin{figure}[tbp]
\centering
\includegraphics[width=0.5\textwidth]{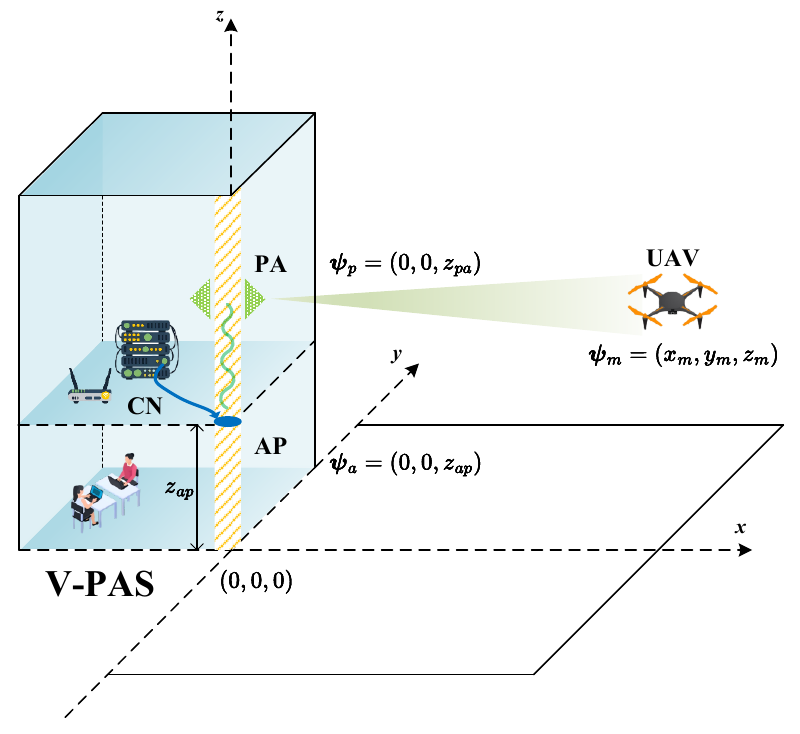}
\caption{Vertical pinching antenna system.}
\label{fig:Model1}
\end{figure}

\section{System Model}
\label{section:2}
\subsection{Deployment Model}
As shown in Fig. \ref{fig:Model1}, we consider a downlink communication scenario aimed at establishing reliable wireless communication between a core network (CN) deployed inside a building and a (ultra-)low-altitude UAV located outside the building.
Specifically, the V-PAS with a vertical dielectric waveguide is mounted on the facade of buildings, extending along the $z$-axis perpendicular to the $x$-$y$ plane with a total height of $D_z$. A 3D Cartesian coordinate system is established with the origin $(0,0,0)$ located at the intersection of the dielectric waveguide and the $x$-$y$ plane. In the proposed V-PAS, a pinching antenna (PA) is deployed along the waveguide and its position is denoted by $\boldsymbol{\psi}_{p} = (0,0,z_{pa})$, where $z_{pa}\in [0,D_z]$ is an adjustable PA height.
\footnote{Existing  PAS implementations employ a dielectric core and cladding for guided propagation, while localized radiating points can be formed using pinching elements. Continuous and discrete PA activation are both possible implementation approaches\bluecite{GPA}.}
The CN equipment is deployed inside the building and connected to the feeding point of the dielectric waveguide via feeder cables. This feeding point is referred to as the AP, which is at the same height as the CN equipment, and its position is denoted by $\boldsymbol{\psi}_{a} = (0,0,z_{ap})$. 
Depending on the floor where the CN equipment is deployed, the AP is located at a corresponding floor-level height within $[0,D_z]$. In the analytical model, $z_{ap}$ is treated as a continuous deployment parameter, while practical AP feeding points are generally restricted to discrete floor-level locations. 
The AP represents a fixed feeding point after installation, while the PA provides a dynamic radiating point along the vertical dielectric waveguide.
The receiving UAV is free to move within a 3D cuboid airspace of dimensions $D_x$, $D_y$, and $D_z$. Its position is denoted by $\boldsymbol{\psi}_m = (x_m, y_m, z_m)$, where $x_m$, $y_m$, and $z_m$ are mutually independent and uniformly distributed over $[0, D_x]$, $[-D_y/2, D_y/2]$, and $[0, D_z]$, respectively.
It should be noted that the uniform UAV-location model is adopted to characterize the spatially averaged performance over the entire 3D service region at an arbitrary observation instant.
In the V-PAS, the signal transmitted by the AP propagates through the dielectric waveguide. 
Through electromagnetic coupling between the waveguide and PA, electromagnetic waves are radiated into free space to establish a short-distance, LoS outdoor link with the UAV. The PA position can be flexibly adjusted along the vertical waveguide to adapt to the 3D mobility of the UAV and improve channel conditions.
\subsection{Channel Model}
According to the spherical wave channel model, the free-space channel coefficient between V-PAS and  UAV is given by
\begin{align}
h_1=\frac{\sqrt{\eta }{{e}}^{-j\frac{2\pi}{\lambda }\left\|\boldsymbol\psi_m-\boldsymbol\psi_{p}\right\|}}{\left\|\boldsymbol\psi_m-\boldsymbol\psi_{p}\right\|},
\label{eq:h1}
\end{align}
where $\eta = \frac{\lambda^2}{16\pi^2}$ denotes the path loss at the reference distance of 1 m, $\lambda = \frac{c}{f_c}$ denotes the free-space wavelength of the signal, $c$ is the speed of light, $f_c$ is the carrier frequency, $j$ is the imaginary unit, and $\left\|\cdot\right\|$ denotes the Euclidean norm.

As the signal propagates through the dielectric waveguide, it interacts with the dielectric core and surrounding materials due to the propagation characteristics of the medium, resulting in a reduction in its phase velocity. This effect is characterized by the effective refractive index $n_{\text{eff}}$, which in turn determines the guided wavelength as $\lambda_g = \lambda/n_{\text{eff}}$. Consequently, the signal undergoes an additional phase shift when traveling from the AP to the PA through the waveguide, which is given by
\begin{align}
h_2={{{e}}^{-j\frac{2\pi}{\lambda_g }\left\|\boldsymbol\psi_{p}-\boldsymbol\psi_{a}\right\|}}.
\label{eq:h2}
\end{align}

According to the solution of Maxwell's equations in lossy media, signals propagating along the dielectric waveguide undergo intrinsic exponential power attenuation, which we characterize by the absorption coefficient $\alpha \in [0, +\infty)$.
\subsection{Signal Model}
Based on the channel coefficients in \textcolor{blue}{\eqref{eq:h1}} and \textcolor{blue}{\eqref{eq:h2}} and the absorption coefficient of the dielectric waveguide, the received signal at UAV in the V-PAS can be expressed as
\begin{align}
y=\sqrt{P_te^{-\alpha\left\|\boldsymbol\psi_{p}-\boldsymbol\psi_{a}\right\| }}h_1h_2s+n,
\label{eq:y}
\end{align}
where $s$ denotes the transmitted signal satisfying $\mathbb{E}\left[|s|^2\right] = 1$, $\mathbb{E}[\cdot]$ denotes the expectation, $n \sim \mathcal{CN}\left(0, \sigma^2\right)$ is the additive white Gaussian noise (AWGN) at UAV, and $P_{t}$ is the transmission power of AP.

Therefore, the signal-to-noise ratio (SNR) of the signal received at UAV in the V-PAS is given by
\begin{align} 
\mathrm{SNR} &= \frac{\eta P_t e^{-\alpha\left\|\boldsymbol{\psi}_p - \boldsymbol{\psi}_a\right\|} \left| e^{-j\left(\frac{2\pi}{\lambda }\left\|\boldsymbol{\psi}_m - \boldsymbol{\psi}_p\right\| + \frac{2\pi}{\lambda_g }\left\|\boldsymbol{\psi}_p - \boldsymbol{\psi}_a\right\|\right)} \right|^2}{\sigma^2 \left\|\boldsymbol{\psi}_m - \boldsymbol{\psi}_p\right\|^2} \notag \\
&\overset{(a)}{=} \frac{\eta \rho _t e^{-\alpha\left\|\boldsymbol{\psi}_p - \boldsymbol{\psi}_a\right\|}}{ \left\|\boldsymbol{\psi}_m - \boldsymbol{\psi}_p\right\|^2},
\label{eq:SNR} 
\end{align}
where step (a) uses the definition $\rho_t = \frac{P_t}{\sigma^2}$ and the equation $|e^{-jx}| = 1$.\footnote{The present analysis focuses on large-scale path loss under the LoS model.
Doppler introduces a time-varying propagation phase and does not directly alter the instantaneous SNR magnitude in the considered model. Nevertheless, practical Doppler effects may increase the burden of synchronization, channel estimation, and PA positioning.}
For a moving UAV with trajectory $\boldsymbol{\psi}_m(t)$, the instantaneous SNR expression remains applicable by replacing the spatial coordinates with their time-varying counterparts.

\textit{Remark 1:}
The single-PA analysis can be extended to the multi-PA case by activating $N$ closely spaced PAs along the dielectric waveguide, with an inter-PA spacing greater than $\lambda/2$ to mitigate mutual coupling\bluecite{MC}.
Let $d_n=\|\boldsymbol{\psi}_m-\boldsymbol{\psi}_{p,n}\|$ and $\ell_n=\|\boldsymbol{\psi}_{p,n}-\boldsymbol{\psi}_a\|$, where $\boldsymbol{\psi}_{p,n}=(0,0,z_{p,n})$ and $n=1,\ldots,N$. 
Taking the first PA as the phase reference, the PA positions can be adjusted such that their composite propagation phases
$\phi_n=\frac{2\pi}{\lambda}d_n+\frac{2\pi}{\lambda_g}\ell_n$
satisfy
\begin{align}
\phi_n-\phi_1=2\pi k_n,\; k_n\in\mathbb{Z},
\end{align}
thereby enabling constructive combining at the UAV.
For closely spaced PAs with $d_n\approx d_1$ and
$e^{-\alpha\ell_n/2}\approx e^{-\alpha\ell_1/2}$,
we further adopt an equal power model with $P_n=P_t/N$,
and the received SNR can be approximated as
\begin{align}
\mathrm{SNR}_N=\frac{\eta}{\sigma^2}\left|\sum_{n=1}^{N}\sqrt{P_n}\frac{e^{-\alpha\ell_n/2}}{d_n}e^{-j\phi_n}\right|^2
\approx N\frac{\eta\rho_t e^{-\alpha\ell_1}}{d_1^2}.
\end{align}
Therefore, an approximately $N$-fold SNR gain can be achieved compared with the corresponding single-PA case.

\textit{Remark 2:} For the V-PAS with an ideal lossless waveguide, the absorption coefficient $\alpha = 0$. When $\alpha = 0$, \textcolor{blue}{\eqref{eq:SNR}} reduces to
\begin{equation}
\mathrm{SNR}_{\text{ideal}} = \frac{\eta \rho_t}{\left\|\boldsymbol{\psi}_m - \boldsymbol{\psi}_p\right\|^2}.
\end{equation}
This expression is solely determined by the free-space path loss and represents the SNR upper bound of V-PAS. In practical deployments, the dielectric waveguide always exhibits non-zero path loss, resulting in an actual SNR lower than the ideal case. This ideal upper bound provides a useful reference for evaluating the performance of V-PAS with a lossy waveguide.

\textit{Remark 3:}
Although this work focuses on a representative UAV for tractable analytical characterization, the proposed V-PAS framework can be naturally extended to multi-UAV scenarios. Under TDMA-based access, the PA can be repositioned according to the scheduled UAV, and the proposed performance analysis can serve as the per-UAV analytical foundation. More advanced schemes, such as WD, WM, WS, and NOMA-based access,  provide promising directions for extending V-PAS toward multi-UAV communication systems.
\section{Performance Analysis}
\label{section:3}
In this section, we analyze the performance of UAV communications achieved by the V-PAS and derive closed-form expressions for outage probability and ergodic rate, respectively, when PA is deployed at the position that minimizes its distance to UAV, i.e., at the same altitude as UAV.
\subsection{Outage Probability}
Deploying the PA at altitude $z_m$ ensures the minimal LoS path loss. However, the communication is subject to outage due to the randomness of the UAV's position.
The outage probability is defined as
\begin{align}
P_{{out}} = \mathrm{Pr}\left( \mathrm{SNR} < \mathrm{SNR}_{\text{thr}} \right),
\label{eq:OP} 
\end{align}
where $\mathrm{SNR}_{\text{thr}}$ is the predefined SNR threshold required for reliable communication. 
An outage occurs when $\mathrm{SNR}$ falls below $\mathrm{SNR}_{\text{thr}}$.
Substituting \textcolor{blue}{\eqref{eq:SNR}} into \textcolor{blue}{\eqref{eq:OP}}, the outage probability of V-PAS with a lossy waveguide can be expressed as
\begin{align}
P_{out}&=\mathrm{Pr}\left(\frac{\eta \rho_te^{-\alpha\left\|\boldsymbol\psi_p-\boldsymbol\psi_a\right\|}}{\left\|\boldsymbol\psi_m-\boldsymbol\psi_p\right\|^2}<\mathrm{SNR_{\text{thr}}}\right), \notag \\
&=\mathrm{Pr}\left(\frac{\eta \rho_te^{-\alpha\left\|z_{pa}-z_{ap}\right\|}}{\left\|{\sqrt{{x_m}^2+{y_m}^2+{(z_m-z_{pa})}^2}}\right\|^2}<\mathrm{SNR_{thr}}\right).
\label{eq:OP1} 
\end{align}

Throughout the following analysis, we consider the common case, i.e., \(D_x > D_y/2\). The complementary case follows the same derivation approach.
Based on \textcolor{blue}{\eqref{eq:OP1}}, the outage probability of UAV communications achieved by the V-PAS is presented in the following proposition.

\textit{Proposition 1:}
The accurate closed-form expressions for the outage probability of V-PAS with a lossy waveguide are shown in Table \ref{tab:OP}, which is at the top of the next page, where $C={\eta \rho _t}/{\mathrm{SNR_{thr}}}$, $H_1=z_{ap}$, $H_2=D_z-z_{ap}$, $T_1=\frac{{D_y}^2}{4}$, $T_2={D_x}^2$, $T_3={D_x}^2+\frac{{D_y}^2}{4}$, $u_n=\frac{1}{\alpha }\ln{\frac{C}{T_n}}$, $m_{n,i}=\min\left ( u_n,H_i\right )$, $n\in \left \{ 1,2,3\right \}$, $i\in \left \{ 1,2\right \}$, $\omega_k=\cos \left ( \frac{2k-1}{2M}\pi\right )$, and $M$ is the accuracy versus complexity parameter.
Moreover, the special function $S_{2}(\cdot)$ involved in Table \ref{tab:OP} is defined as
\begin{align}
S_2(k)=&2\left ( \int_{0}^{\sqrt{k-\frac{{D_y}^2}{4}}}\frac{D_y}{2}\mathrm{d}x+\int_{\sqrt{k-\frac{{D_y}^2}{4}}}^{\sqrt{k}}\sqrt{k-x^2}\mathrm{d}x\right ) \notag \\
=&\frac{D_y}{2}\sqrt{k-\frac{{D_y}^2}{4}}+\frac{\pi k}{2}-k\arccos \left ( \frac{D_y}{2\sqrt{k}}\right ),
\label{eq:S2_k}
\end{align}
and the special function $S_{3}(\cdot)$ in Table \ref{tab:OP} is defined as
\begin{align}
S_3(k) = &2\left ( \int_{0}^{\sqrt{k-\frac{{D_y}^2}{4}}}\frac{D_y}{2}\mathrm{d}x+\int_{\sqrt{k-\frac{{D_y}^2}{4}}}^{D_x}\sqrt{k-x^2}\mathrm{d}x\right ) \notag \\
=&\frac{D_y}{2}\sqrt{k-\frac{{D_y}^2}{4}}+D_x\sqrt{k-{D_x}^2}\notag \\
&+k\left [ \arcsin\left ( \frac{D_x}{\sqrt{k}}\right ) -\arccos \left ( \frac{D_y}{2\sqrt{k}}\right )\right ].
\label{eq:S3_k}
\end{align}

\begin{IEEEproof}
See Appendix A.
\end{IEEEproof}


\begin{table*}[!t]
\centering
\caption{The conditions and accurate closed-form expressions for outage probability $P_{out}$}
\begin{tabular}{l|l}
\hline\hline  
\textbf{Conditions} & \textbf{The Closed-Form Expressions} \\
\hline\hline  
$\displaystyle 0<C\leq \frac{{D_y}^2}{4}$ & $\displaystyle 1-\frac{\pi C}{2\alpha D_xD_yD_z}\left [ 2-e^{-\alpha z_{ap}}-e^{-\alpha \left ( D_{z} - z_{ap}\right )}\right ]$ \\
\hline  
$\displaystyle \frac{{D_y}^2}{4}<C\leq {D_x}^2$ & $\displaystyle 1-\frac{1}{D_xD_yD_z}\displaystyle\sum_{i=1}^{2}\left [ \frac{m_{1,i}\pi}{2M}\displaystyle\sum_{k=1}^{M}S_{2}\!\left(C e^{-\alpha\left(\frac{m_{1,i}}{2}\omega_k+\frac{m_{1,i}}{2}\right)}\right)\sqrt{1-{\omega_k}^2}+\frac{\pi C}{2\alpha }\left (e^{-\alpha {m_{1,i}}}-e^{-\alpha {H_{i}}} \right )\right ]$ \\
\hline  
$\displaystyle {D_x}^2<C\leq {D_x}^2+\frac{{D_y}^2}{4}$ & 
$\displaystyle 
\begin{aligned}[t]
&1-\frac{1}{D_xD_yD_z}\sum_{i=1}^{2}\bigg[ 
     \frac{m_{2,i}\pi}{2M}\sum_{k=1}^{M}S_{3}\!\left(C e^{-\alpha\left(\frac{m_{2,i}}{2}\omega_k+\frac{m_{2,i}}{2}\right)}\right)\sqrt{1-{\omega_k}^2} \\
    & + \frac{\left ( m_{1,i}-m_{2,i}\right )\pi}{2M}\sum_{k=1}^{M}S_{2}\!\left(C e^{-\alpha\left(\frac{ m_{1,i}-m_{2,i}}{2}\omega_k+\frac{m_{1,i}+m_{2,i}}{2}\right)}\right)\sqrt{1-{\omega_k}^2}  + \frac{\pi C}{2\alpha }\left (e^{-\alpha {m_{1,i}}}-e^{-\alpha {H_{i}}} \right )\bigg]
\end{aligned}$ \\
\hline  
$\displaystyle C>{D_x}^2+\frac{{D_y}^2}{4}$ & $\displaystyle 
\begin{aligned}[t]
&1-\frac{1}{D_xD_yD_z}\sum_{i=1}^{2}\bigg [D_xD_ym_{3,i}+\frac{\left ( m_{2,i}-m_{3,i}\right )\pi}{2M}\sum_{k=1}^{M}S_{3}\!\left(C e^{-\alpha\left(\frac{m_{2,i}-m_{3,i}}{2}\omega_k+\frac{m_{2,i}+m_{3,i}}{2}\right)}\right)\sqrt{1-{\omega_k}^2}\\
& + \frac{\left ( m_{1,i}-m_{2,i}\right )\pi}{2M}\sum_{k=1}^{M}S_{2}\!\left(C e^{-\alpha\left(\frac{ m_{1,i}-m_{2,i}}{2}\omega_k+\frac{m_{1,i}+m_{2,i}}{2}\right)}\right)\sqrt{1-{\omega_k}^2}+\frac{\pi C}{2\alpha }\left (e^{-\alpha {m_{1,i}}}-e^{-\alpha {H_{i}}} \right )\bigg ]
\end{aligned}$ \\
\hline\hline  
\end{tabular}
\label{tab:OP}
\end{table*}

To gain more insights, we investigate the asymptotic behavior of V-PAS as the transmission power \(P_t\) increases.

\textit{Corollary 1:} The asymptotic outage probability of V-PAS is given by
\begin{equation}
\label{eq:OPasy}        
P_{out,asy}=\begin{dcases} \frac{q\left(D_x^2 + D_y^2/4\right)^2}{6\,\alpha\,D_z\,D_x^2 D_y^2}
\left[\ln\!\left(\frac{P_{\mathrm{th}}}{P_t}\right)\right]^3, &P_t \to P_{\mathrm{th}}^-,
\\ 
0, &P_t\ge{P_\mathrm{th}},
\end{dcases}
\end{equation}
where \(P_{\mathrm{th}} = \frac{\sigma^2  \mathrm{SNR_{thr}}}{\eta} \left( D_x^2 + \frac{D_y^2}{4} \right) e^{\alpha \max(z_{ap},\, D_z-z_{ap})}
\), \(q = 2\) for \(z_{ap} = D_z/2\), and \(q = 1\) otherwise.

\begin{IEEEproof}
See Appendix B.
\end{IEEEproof}

\textit{Remark 4:}
Within the confined UAV operational area, V-PAS exhibits asymptotic behavior different from that in unbounded scenarios.
In particular, there exists a finite transmission power threshold \(P_{\mathrm{th}}\) such that when \(P_t \ge P_{\mathrm{th}}\), even the farthest UAV within the area satisfies the SNR requirement, resulting in zero outage probability.
As $P_t$ approaches $P_{\mathrm{th}}$ from below, the outage probability decays as \(\bigl[\ln(P_{\mathrm{th}}/P_t)\bigr]^3\), a rate faster than that of the diversity gains  in unbounded scenarios.
This result provides a clear critical power reference for achieving zero outage and highlights the nature of finite-area deployment.
\subsection{Ergodic Rate}
In addition to the outage probability, the ergodic rate is another key performance metric to evaluate the transmission performance of wireless communication systems. The ergodic rate is defined as the statistical expectation of the instantaneous achievable rate, reflecting the long-term average information transmission capability of the system under random channel conditions. The ergodic rate is mathematically defined as 
\begin{equation} 
\label{eq:Rp} 
R_p = \mathbb{E}\left[ \log_2\left( 1 + \mathrm{SNR} \right) \right],
\end{equation}
where $\mathbb{E}[\cdot]$ denotes the expectation. Then, the ergodic rate of UAV communications achieved by the V-PAS is presented as follows.

\textit{Proposition 2:}
The accurate closed-form expression for the ergodic rate of V-PAS with a lossy waveguide is given by
\begin{align}
R_p = &\frac{1}{\ln 2} \displaystyle\sum_{i=1}^{3} \biggl[ \frac{\pi a_i}{M} \displaystyle\sum_{k=1}^{M} \sqrt{1-\omega_k^2} \notag \\
&\times Q\left( a_i\omega_k + b_i \right) f_{R,i}\left( a_i\omega_k + b_i \right) \biggr],
\label{eq:Rp1}
\end{align}
where $a_1=b_1=\frac{D_y}{4}$, $a_2=\frac{D_x-\frac{D_y}{2}}{2}$, $b_2=\frac{D_x+\frac{D_y}{2}}{2}$, $a_3=\frac{\sqrt{{D_x}^2+\frac{{D_y}^2}{4}}-D_x}{2}$, $b_3=\frac{\sqrt{{D_x}^2+\frac{{D_y}^2}{4}}+D_x}{2}$, $Q\left ( r\right )$ is given by
\begin{align}
Q\left ( r\right ) =& \frac{1}{\alpha D_z}\biggl[ \operatorname{Li}_2\left ( -\frac{\eta \rho_te^{-\alpha z_{ap}}}{r^2}\right )\notag \\
&+\operatorname{Li}_2\left ( -\frac{\eta \rho_te^{-\alpha \left ( D_z-z_{ap}\right )}}{r^2}\right )-2 \operatorname{Li}_2\left ( -\frac{\eta \rho_t}{r^2}\right )\biggr ],
\label{eq:Q} 
\end{align}
$\operatorname{Li}_2(x) = -\int_0^x \frac{\ln(1-u)}{u}\mathrm{d}u$ is the dilogarithm function in \textcolor{blue}{[}\citenew{toolbook}, Eq. (6.254.1)\textcolor{blue}{]}, and $f_{R,i}\left( r \right)$
is given by 
\begin{equation}
\label{eq:f_R}        
f_{R,i}\left( r \right) =\begin{dcases}\frac{\pi r}{D_xD_y}, &i=1,
\\ 
\frac{\pi r-2r\arccos \left ( \frac{D_y}{2r}\right )}{D_xD_y}, &i=2,
\\
\frac{2r\left ( \arcsin \frac{D_x}{r}-\arccos  \frac{D_y}{2r}\right )}{D_xD_y}, &i=3,
\\
0, &i=4.
\end{dcases}
\end{equation}

\begin{IEEEproof}
See Appendix C.
\end{IEEEproof}

\textit{Corollary 2:} As the transmission power \(P_t \to \infty\), the asymptotic ergodic rate of V-PAS is given by
\begin{align}
R_{p,asy} =& \log_2P_t + \log_2\eta -\log_2\sigma^2- \frac{2\mathbb{E}\left[\ln\left(\sqrt{x_m^2+y_m^2}\right)\right]}{\ln 2} \notag\\
&- \frac{\alpha}{2 D_z \ln 2}\left[z_{ap}^2 + (D_z-z_{ap})^2\right],      
\label{eq:Rp_asy}  
\end{align}
where \(\mathbb{E}\left[\ln\left(\sqrt{x_m^2+y_m^2}\right)\right]\) denotes the expected value of the logarithm of the horizontal distance between UAV and AP, which is a  constant determined by  \(D_x\) and \(D_y\).
Specifically, \(\mathbb{E}\left[\ln\left(\sqrt{x_m^2+y_m^2}\right)\right]\)  is given by
\begin{align}
\mathbb{E}\left[\ln\left(\sqrt{x_m^2+y_m^2}\right)\right] =& \frac{1}{2}\ln\left(D_x^2 + \frac{D_y^2}{4}\right) - \frac{3}{2} + \frac{\pi D_x}{2 D_y} \notag\\
&- \frac{D_x^2 - D_y^2/4}{D_x D_y} \arctan\left(\frac{2D_x}{D_y}\right). 
\label{eq:Exy}  
\end{align}

\begin{IEEEproof}
See Appendix D.
\end{IEEEproof}

\textit{Remark 5:}
Within the confined UAV operational area, ergodic rate increases logarithmically with $\log_2 P_t$ as $P_t$ tends to infinity. The constant offset of ergodic rate is jointly determined by path loss, AWGN, absorption coefficient, UAV operational area, and AP height. At high SNR, these system constants merely shift the rate versus log-power curve vertically.

\section{Pinching Multiplicative Path Loss and Deployment Optimality}
\label{section:4}
This section first formally defines the concept of PMPL, which captures the unique cascaded attenuation characteristic of V-PAS. Numerical results are presented to quantify the performance advantage region of V-PAS and verify its adaptability to building heights. 
Subsequently, we investigate the midpoint symmetry of V-PAS with respect to AP height and characterize the optimal PA position under waveguide loss, providing insights into the deployment of both AP and PA.
\subsection{Pinching Multiplicative Path Loss}
In the practical scenario where the absorption coefficient is non-zero, the path loss of V-PAS consists of two cascaded components. The first component is 
the dielectric waveguide loss from AP to PA, i.e., $e^{\alpha|z_m-z_{ap}|}$, and the second is 
the free-space path loss from PA to UAV, i.e., $\frac{x_m^2+y_m^2}{\eta}$. 
We define the \textit{pinching multiplicative path loss (PMPL)} as the cascaded multiplicative attenuation of the waveguide and free-space path losses, which is given by
\begin{align}
L_{\mathrm{mul}}=\frac{\left ( {x_m}^2+{y_m}^2\right )e^{\alpha\left|z_{m}-z_{ap}\right|}}{\eta}.
\label{eq:PMPL}
\end{align}
For clarity, let $r=\sqrt{x_m^2+y_m^2}$ and $h=|z_m-z_{ap}|$ denote the horizontal and vertical distances between the AP and UAV, respectively. Then, the PMPL can be rewritten as 
$L_{\rm mul}=r^2e^{\alpha h}/{\eta}.$

As a performance benchmark, we consider a conventional path-loss model in which AP communicates directly with UAV via a LoS link. The corresponding path loss is the 3D free-space path loss, 
which is given by 
\begin{align}
L_{\mathrm{dir}}=\frac{x_m^2+y_m^2+(z_m-z_{ap})^2}{\eta}=\frac{r^2+h^2}{\eta}.
\label{eq:L_dir}
\end{align}

The key difference between PMPL and the conventional path loss lies in how the vertical distance contributes to the overall path loss. In the conventional path loss, both the horizontal distance $r$ and vertical distance $h$ contribute to the free-space propagation distance. In V-PAS, however, the vertical distance affects PMPL through the waveguide loss factor $e^{\alpha h}$, and only the horizontal distance determines the PA-UAV free-space path loss. 
Accordingly, the conventional path loss is governed by the complete 3D Euclidean free-space distance, while PMPL is jointly governed by the PA-UAV free-space path loss and the AP-PA waveguide loss.
This difference becomes particularly clear for the ideal lossless waveguide with $\alpha=0$, where $L_{\rm mul}=r^2/\eta$ becomes independent of the vertical distance $h$, while the conventional path loss remains unchanged.

\textit{Remark 6:}
Comparing \textcolor{blue}{\eqref{eq:PMPL}} and \textcolor{blue}{\eqref{eq:L_dir}}, V-PAS achieves a lower path loss than the benchmark when $L_{\mathrm{mul}}<L_{\mathrm{dir}}$.
This condition can be rewritten as
\begin{align}
r<r_{\mathrm{th}}(h,\alpha)=\frac{h}{\sqrt{e^{\alpha h}-1}},
\label{r_th}
\end{align}
where $h>0$.
Therefore, $r_{\mathrm{th}}(h,\alpha)$ directly characterizes the horizontal advantage region of V-PAS for a given vertical distance and absorption coefficient. A smaller $\alpha$ results in a larger $r_{\mathrm{th}}(h,\alpha)$, indicating that the use of a lower-loss dielectric waveguide enlarges the horizontal region in which V-PAS has a path-loss advantage over the benchmark.
A quantitative comparison between PMPL and direct path loss is presented below. The results reveal distinct performance behaviors across different horizontal and vertical distance ranges.

\begin{figure}[t]
\centering
\includegraphics[width=\columnwidth]{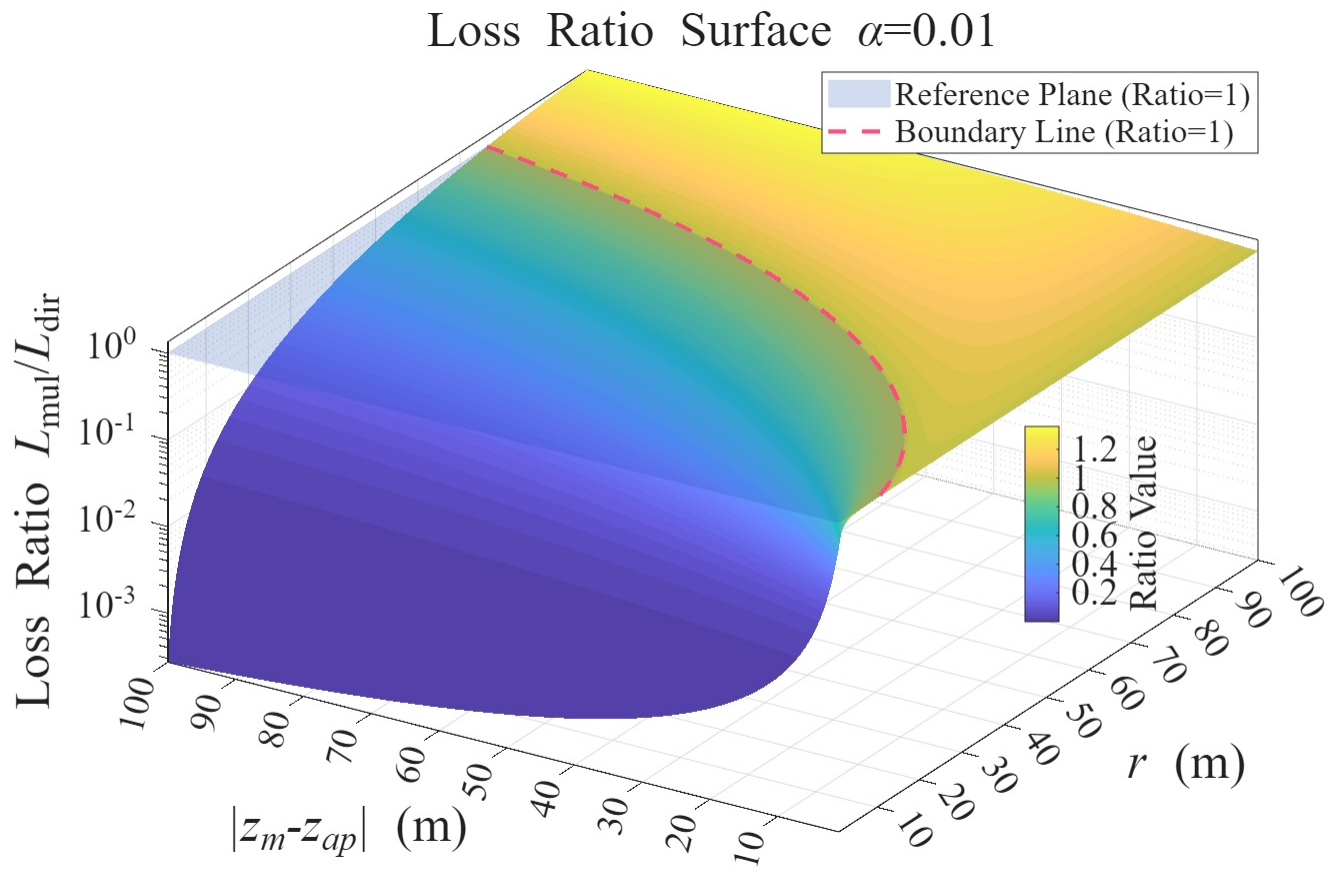}
\caption{Ratio of PMPL to conventional path loss versus $|z_m-z_{ap}|$ and $r$, where $\alpha=0.01$.}
\label{fig:loss_ratio}
\end{figure}

Fig. \ref{fig:loss_ratio} shows the ratio of PMPL to conventional path loss as a function of $|z_m-z_{ap}|$ and $r$, where $\alpha=0.01$ Np/m.
A ratio smaller than 1 indicates that PMPL is smaller than the direct path loss and thus V-PAS enjoys a performance advantage, whereas a ratio larger than 1 implies the direct path loss is superior.
From this figure, key conclusions can be drawn.
1)
When the horizontal distance is small, i.e., $r<r_{\mathrm{th}}(1,0.01)\approx 9.96$ m, the loss ratio remains far below 1 regardless of the vertical distance ranging from 0 m to 100 m, and the minimum value reaches the order of $10^{-3}$. This result clearly demonstrates the core advantage of V-PAS in “last meter" communication scenarios. 
More importantly, within this region the vertical distance has almost no effect on PMPL, which confirms that the proposed V-PAS is highly adaptive to building heights and can achieve significant performance gains for both low‑rise and high‑rise buildings.
2)
When the horizontal distance falls in the intermediate range, i.e., $r_{\mathrm{th}}(1,0.01)<r<r_{\mathrm{th}}(100,0.01)$, the relative superiority between V-PAS and the benchmark exhibits a clear crossover behavior. For small vertical distances, PMPL exceeds the direct path loss. However, as the vertical distance increases, the ratio gradually decreases and falls below 1. This indicates that under intermediate horizontal distances, a larger vertical distance is actually beneficial for V-PAS to exploit its performance advantage.
3)
When the horizontal distance is large, i.e., $r>r_{\mathrm{th}}(100,0.01)\approx76.3$ m, the loss ratio stays above 1 regardless of the vertical distance. This implies that in long‑distance communication scenarios, PMPL introduces an additional performance penalty and V‑PAS is no longer applicable.
4)
The red dashed curve in Fig. \ref{fig:loss_ratio} corresponds to the analytical boundary $r_{\mathrm{th}}(h,\alpha)$ and marks the contour where the ratio is equal to 1, which divides the entire parameter space into the V-PAS-advantage region and the direct-communication-advantage region. This contour expands outward as the vertical distance increases, further verifying that a larger vertical distance helps broaden the applicability of V-PAS. 
These results provide explicit quantitative guidance for practical system deployment, i.e., V-PAS is best suited for “last meter" communication scenarios rather than “last hundred meters."

\begin{figure}[t]
\centering
\subfigure[\scriptsize $L_{\mathrm{dir}}$ contours.]{
    \includegraphics[width=0.465\columnwidth]{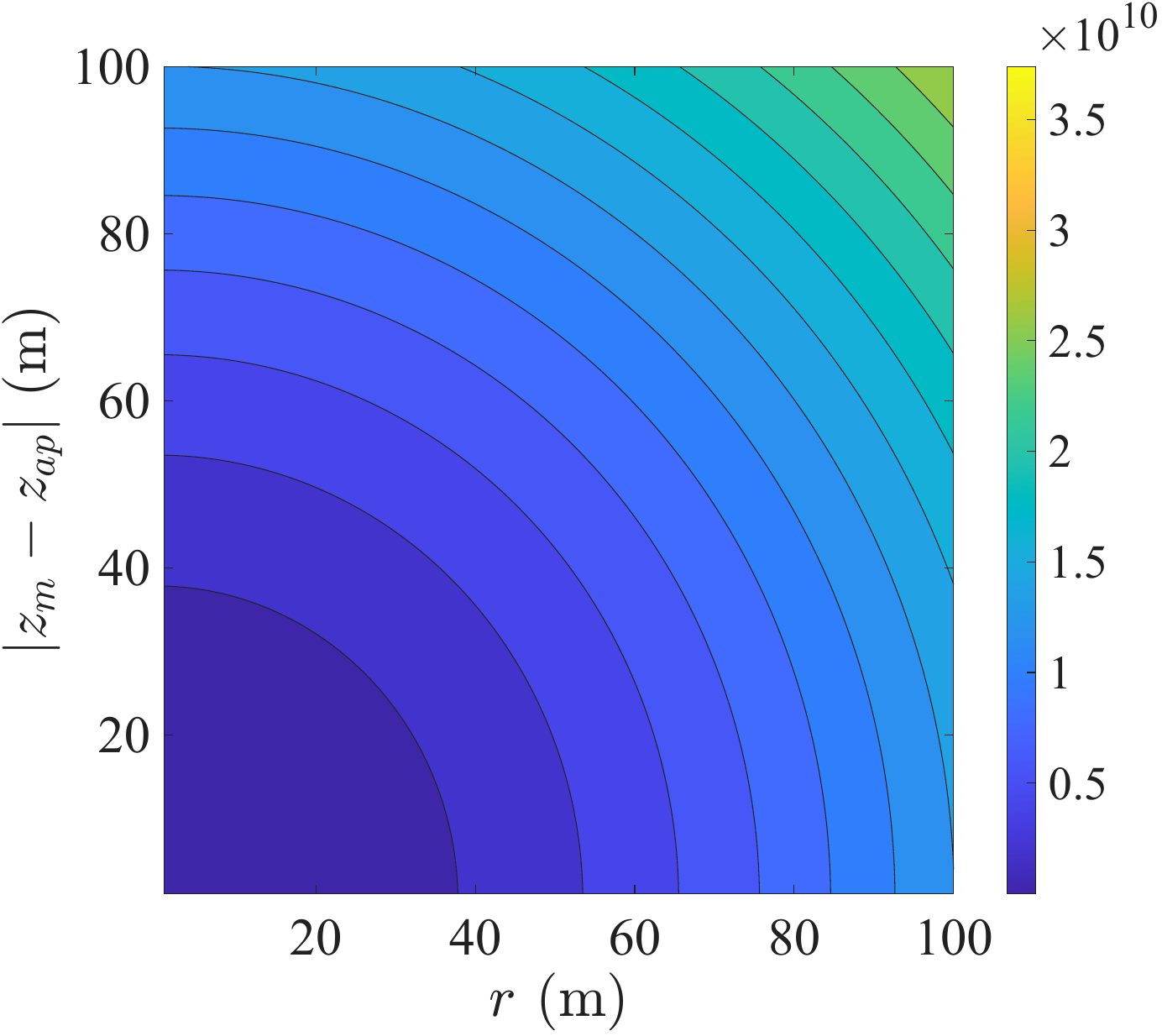}
    \label{fig:L_dir}}
\subfigure[\scriptsize $L_{\mathrm{mul}}$ contours.]{
    \includegraphics[width=0.465\columnwidth]{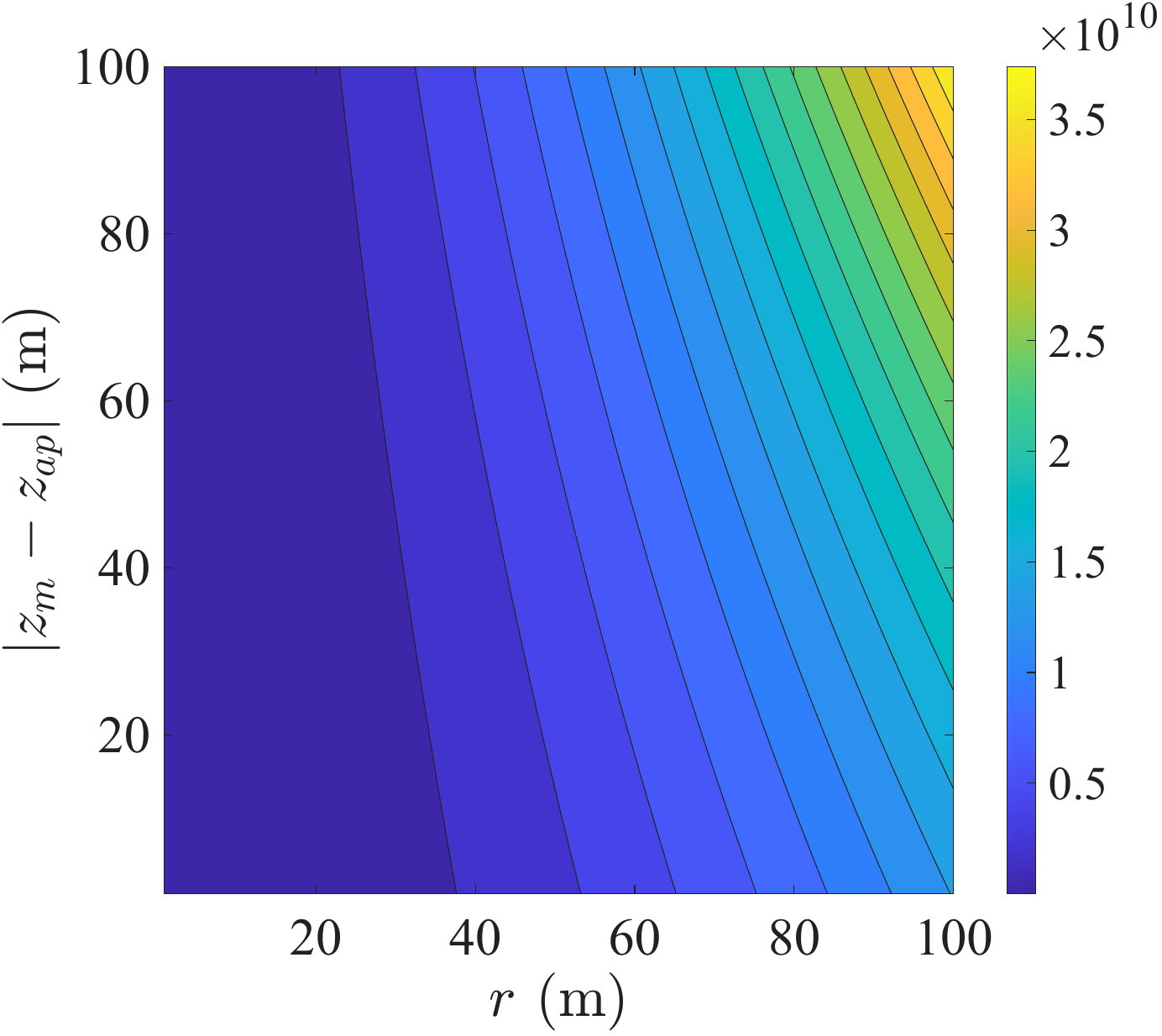}
    \label{fig:L_mul}}
\caption{Contours of PMPL and conventional path loss versus $|z_m-z_{ap}|$ and $r$, where $\alpha=0.01$.}
\label{fig:L_contours}
\end{figure}

To further explain the observed insensitivity to vertical distance, 
Fig. \ref{fig:L_contours} shows the contours of PMPL and conventional path loss versus $|z_m-z_{ap}|$ and $r$, where $\alpha=0.01$.
Key conclusions from this figure are summarized as follows.
1)
In Fig. \ref{fig:L_dir}, the contours of the direct path loss exhibit standard concentric circular arcs, a typical characteristic of the spherical wave channel model. This indicates that the direct path loss is equally sensitive to variations in horizontal and vertical distances, and its value depends solely on the Euclidean distance between AP and UAV.
2)
In contrast, the PMPL contours in Fig. \ref{fig:L_mul} appear as nearly parallel inclined lines with very steep slopes. This indicates that PMPL is much more sensitive to changes in horizontal distance than to changes in vertical distance, being relatively insensitive to the latter. 
3)
Notably, when the horizontal distance $r < 35$ m, the PMPL contours become almost vertical, implying that variations in vertical distance have virtually no effect on PMPL. This result is fully consistent with the loss ratio analysis in Fig. \ref{fig:loss_ratio}, further revealing the core advantage of V-PAS. 
This contrast arises from the different effects of the vertical distance on the two path-loss models. In $L_{\rm dir}$, the vertical distance directly contributes to the 3D free-space propagation distance, while in $L_{\rm mul}$, the vertical distance affects PMPL through the waveguide loss factor.
Due to the inherent insensitivity of PMPL to vertical height differences, the proposed V-PAS is highly adaptive to building heights. Both low-rise and high-rise buildings can achieve stable and excellent transmission performance when the horizontal distance is small.

\subsection{Midpoint Symmetry of AP Deployment}
Based on the accurate closed-form expressions from Table \ref{tab:OP} and \textcolor{blue}{\eqref{eq:Rp1}}, one can observe that both outage probability and ergodic rate, as functions of AP height $z_{{ap}}$, are strictly symmetric with respect to $z_{{ap}} = D_z/2$. This symmetry originates from the uniform distribution of the UAV's position. In particular, if we replace $z_{{ap}}$ with $D_z - z_{{ap}}$ and simultaneously map the vertical position of UAV from $z_m$ to $D_z - z_m$, the absolute value or square of the vertical distance $z_{{ap}} - z_m$ remains invariant. Consequently, the statistical averages of the performance metrics under these two heights are identical.

\textit{Proposition 3:}
The physical interpretation is that when the AP is located at the midpoint of the vertical deployment range, the vertical position of uniformly distributed UAV is on average closest to the AP. This results in the minimum outage probability and the maximum ergodic rate at $z_{{ap}} = D_z/2$, thus providing a clear guideline for optimal AP deployment. 
In a practical multi-story building, the CN equipment and the waveguide feeding point AP are typically selected from a finite set of feasible floor levels according to the existing network infrastructure. Therefore, the feasible floor closest to the midpoint of the vertical service range should be preferred for locating the CN and AP.
\begin{IEEEproof}
The symmetry property and the performance optimality at $z_{{ap}} = D_z/2$ are proven as follows.
Examination of \textcolor{blue}{\eqref{eq:Rp1}} and \textcolor{blue}{\eqref{eq:Q}} reveals that only the first two terms of \textcolor{blue}{\eqref{eq:Q}} depend on $z_{ap}$. And $f(z)$ is defined as
\begin{align}
f(z)=\operatorname{Li}_2\left(-A e^{-\alpha z}\right)+\operatorname{Li}_2\left(-A e^{-\alpha (D_z-z)}\right),
\label{eq:fh}
\end{align}
where $z\in[0,D_z]$, $A=\eta\rho_t/r^2>0$, $\alpha>0$, and $D_z>0$.

Replacing $z$ by $D_z-z$ swaps the two terms, hence $f(D_z-z)=f(z)$. Thus, $f(z)$ is symmetric about $z=D_z/2$.

We set $x=z-D_z/2$ and $x\in[-D_z/2,D_z/2]$. Then \textcolor{blue}{\eqref{eq:fh}} can be rewritten as
\begin{align}
f(z)=\operatorname{Li}_2\left(-B e^{-\alpha x}\right)+\operatorname{Li}_2\left(-B e^{\alpha x}\right)\triangleq  g(x),
\label{eq:gx}
\end{align}
where $B=A e^{-\alpha D_z/2}>0$. The first derivative of $g(x)$ is given by
\begin{align}
g'(x) =& \alpha \ln(1+B e^{-\alpha x}) - \alpha \ln(1+B e^{\alpha x})\notag\\
=& \alpha \ln\!\left(\frac{1+B e^{-\alpha x}}{1+B e^{\alpha x}}\right).
\label{eq:gx'}
\end{align}

At $x=0$, we have $g'(0)=0$. For $x>0$, note that $e^{\alpha x}>e^{-\alpha x}$, which implies $\ln(\cdot)<0$ and hence $g'(x)<0$. Conversely, $g'(x)>0$ for $x<0$. It follows that $g(x)$ is strictly increasing on $[-D_z/2,0]$ and decreasing on $[0,D_z/2]$, reaching its unique maximum at $x=0$. As a result, $Q(r)$ is symmetric about $z_{ap}=D_z/2$ and attains its maximum at this point. Simulation results presented in Figs. \ref{fig:OP_hap_Dxy50_Ps12}, \ref{fig:OP_hap_Dxy100_Ps16}, and \ref{fig:AVG_hap_Dxy100_Ps16} in the next section further corroborate this symmetry property.
\end{IEEEproof}
\subsection{Optimal Position of Pinching Antenna}
In the previous analysis, the PA is placed at the same altitude as the UAV, which minimizes the free-space propagation distance between the PA and UAV. Although this placement is optimal for an ideal lossless waveguide, it does not necessarily maximize the received SNR in the presence of waveguide loss. Specifically, moving the PA toward AP shortens the propagation distance inside the lossy waveguide, while increasing the free-space propagation distance between the PA and UAV. Therefore, the optimal PA position is determined by the trade-off between the waveguide loss and the free-space path loss. To characterize this trade-off, the following proposition provides the optimal PA position that maximizes the received SNR.

\textit{Proposition 4:}
For a UAV located at $\boldsymbol{\psi}_m=(x_m,y_m,z_m)$,
with $r=\sqrt{x_m^2+y_m^2}>0$ and $h=|z_m-z_{ap}|$,
the received SNR is maximized when the PA is located at
\begin{align}
z_{pa}^{*}
=z_{ap}+\operatorname{sgn}(z_m-z_{ap})l^{*},
\label{eq:z_pa*}
\end{align}
where, for $\alpha > 0$, $\operatorname{sgn}(\cdot)$ is the sign function, $l^{*}$ is given by
\begin{align}
l^{*}=\begin{cases}l_o,& 0<\alpha r<1,\; l_o>0,\; \Psi(l_o)>\Psi(0),\\
0,& \text{otherwise},
\end{cases}
\label{eq:l*}
\end{align}
with $l_o=h-\frac{1-\sqrt{1-\alpha^2r^2}}{\alpha}$, and $\Psi(l)=\frac{e^{-\alpha l}}{r^2+(h-l)^2}$.
For the lossless case $\alpha=0$, the optimal PA position
reduces to $l^*=h$, i.e., $z_{pa}^{*}=z_m$.

\begin{IEEEproof}
The optimal PA cannot lie outside the vertical interval between the AP and UAV. Otherwise, moving the PA toward this interval simultaneously reduces both waveguide loss and the free-space path loss, thereby increasing the received SNR.
Therefore, $\min(z_m,z_{ap})\leq z_{pa}^{*}\leq\max(z_m,z_{ap})$.

We define $l=|z_{pa}-z_{ap}|$, $0\leq l\leq h$, as the propagation distance from the AP to PA along the vertical waveguide. Accordingly, the vertical distance between the PA and UAV is $h-l$.
Maximizing the received SNR is equivalent to maximizing
\begin{align}
\Psi(l)=\frac{e^{-\alpha l}}{r^2+(h-l)^2}.
\label{eq:Psi}
\end{align}
The first derivative of $\Psi(l)$ is given by
\begin{align}
\Psi'(l)=\frac{e^{-\alpha l}\left[-\alpha\left(r^2+(h-l)^2\right)+2(h-l)\right]}{\left[r^2+(h-l)^2\right]^2}.
\label{eq:Psi'}
\end{align}
Since $e^{-\alpha l}>0$ and
$\left[r^2+(h-l)^2\right]^2>0$, the critical points of $\Psi(l)$ are
determined by
\begin{align}
\alpha(h-l)^2-2(h-l)+\alpha r^2=0.
\label{eq:real_points}
\end{align}
Considering that $0 <\alpha r<1$, the roots of \textcolor{blue}{\eqref{eq:real_points}} are given by
\begin{align}
l_{\pm}=h-\frac{1\pm\sqrt{1-\alpha^2r^2}}{\alpha}.
\label{eq:l+-}
\end{align}
To determine the nature of the critical points, $\Psi'(l)$ can be rewritten as $\Psi'(l)=\Psi(l)\Phi(l)$, where $\Phi(l)=-\alpha+\frac{2(h-l)}{r^2+(h-l)^2}$ and $\Phi'(l)=
\frac{2\left[(h-l)^2-r^2\right]}
{\left[r^2+(h-l)^2\right]^2}$.
Since $\Psi(l)>0$, an interior critical point satisfying
$\Psi'(l)=0$ equivalently satisfies $\Phi(l)=0$, and the second derivative of $\Psi(l)$ is given by $\Psi''(l)=\Psi(l)\Phi'(l)$.
Since $({h-l_-})({h-l_+})=r^2$ and $h-l_-<h-l_+$, we have $h-l_-<r<h-l_+$. Thus, $\Phi'(l)<0$ at $l=l_-$ and $\Phi'(l)>0$ at $l=l_+$, and $l_-$ corresponds to a local maximum and $l_+$ to a local minimum. 
The only possible interior maximizer is
\begin{align}
l_o=l_-=h-\frac{1-\sqrt{1-\alpha^2r^2}}{\alpha}.
\label{eq:lo}
\end{align}
For $0<\alpha r<1$, the candidate
$l_o=l_-$ is feasible whenever $l_o>0$. As $l_o$ is the
only interior local maximum and $l=h$ cannot be a maximizer since $\Psi'(h)<0$, the global maximizer is determined
by comparing it with the boundary point $l=0$.
For $\alpha r\geq1$, $\Psi(l)$ is nonincreasing over
$0\leq l\leq h$, and $l^*=0$. For the lossless case
$\alpha=0$, $\Psi(l)$ increases with $l$, giving $l^*=h$.
The proof of Proposition 4 is completed.
\end{IEEEproof}

\textit{Remark 7:} 
Proposition 4 provides two insights into the practical deployment of V-PAS. First, since the absorption coefficient $\alpha$ is very small in practice, the condition $\alpha r\ll1$ generally holds for a moderate communication distance. In this case, using $\sqrt{1-\alpha^2r^2}\approx1-\alpha^2r^2/2$, we have $l_o\approx h-\alpha r^2/2\approx h$. Consequently, the optimal PA altitude can be approximated by the UAV altitude, which reduces the computational complexity associated with PA position optimization. Second, when $\alpha r\geq1$, Proposition 4 shows that the optimal PA is located at the same altitude as the AP, and the signal no longer needs to propagate along the waveguide. Therefore, for a given nonzero absorption coefficient, this result indicates a potential limitation of PAS for ultra-long-distance communications.

\section{Simulation Results and Discussion}
\label{section:5}
In this section, we evaluate the performance of UAV communications achieved by the V-PAS as well as the accuracy and validity of the derived closed-form expressions.
Without loss of generality, we assume that $D_z=100$ m, $f_c=28$ GHz, $n_{\text{eff}}=1.4$, $\sigma^2=-90$ dBm, $\mathrm{SNR}_{\text{thr}}=5$, and $M=400$. 
Unless otherwise stated, the AP height is set as $z_{ap}=10$ m and the absorption coefficient $\alpha$ takes values of $0.003$, $0.005$, and $0.01$ Np/m.
\footnote{
The carrier frequency $f_c=28$ GHz is adopted in representative PAS studies and is also covered by 5G NR band n257\bluecite{5G}.
The absorption coefficient \(\alpha\) is in nepers per meter (Np/m). The conversion relation between \(\mathrm{Np/m}\) and \(\mathrm{dB/m}\) is given by \(\alpha_{\mathrm{dB/m}}=\frac{20}{\ln10}\alpha_{\mathrm{Np/m}}\approx 8.686\alpha_{\mathrm{Np/m}}\)\bluecite{GPA}. 
}
Moreover, we also consider an ideal lossless dielectric waveguide to provide a performance upper bound. To verify the accuracy of these expressions, Monte Carlo simulations are conducted over $10^6$ independent channel realizations. For performance comparison, we consider the conventional path-loss model where AP communicates directly with UAV without deploying any PA as the benchmark.

\begin{figure}[t]
\centering
\subfigure[\scriptsize $D_x=D_y=50$ m.]{
    \includegraphics[width=0.465\columnwidth]{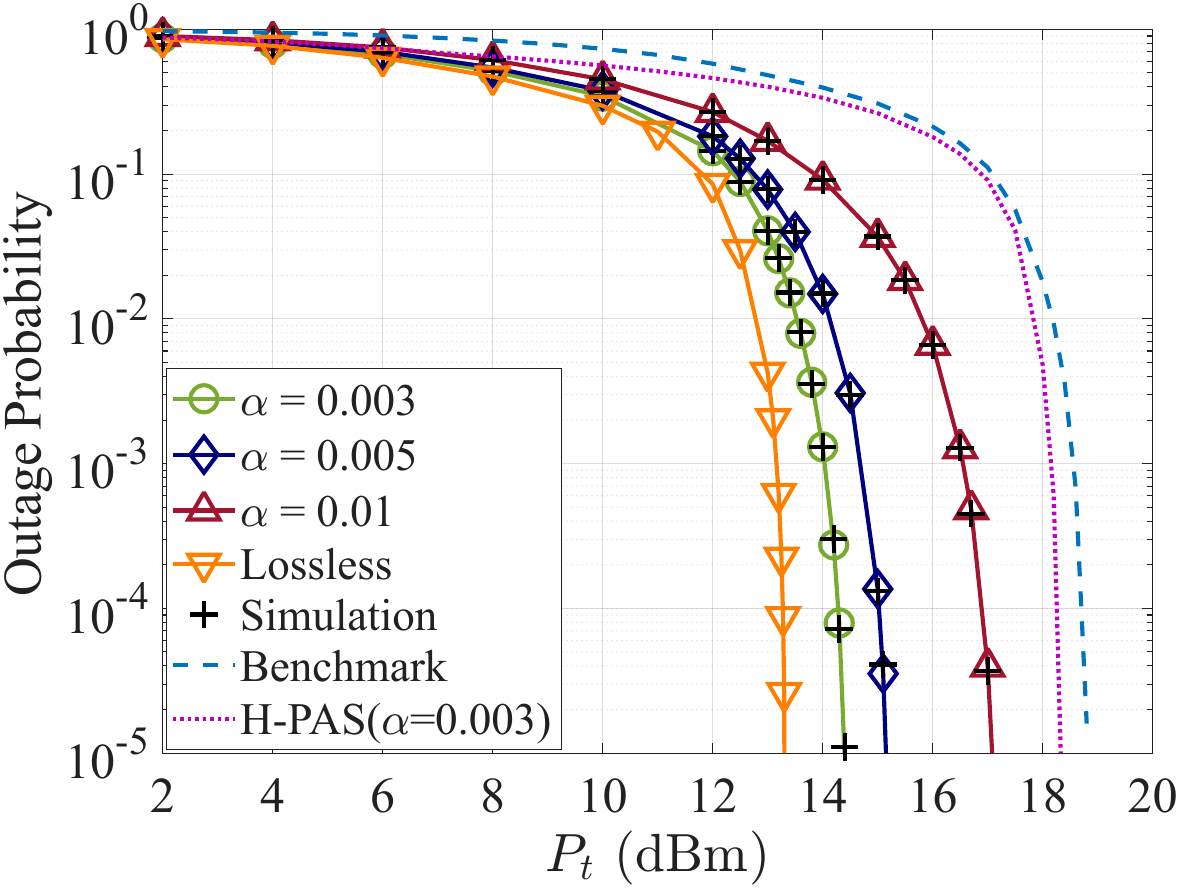}
    \label{fig:OP_Ps_Dxy50}}
\subfigure[\scriptsize $D_x=D_y=100$ m.]{
    \includegraphics[width=0.465\columnwidth]{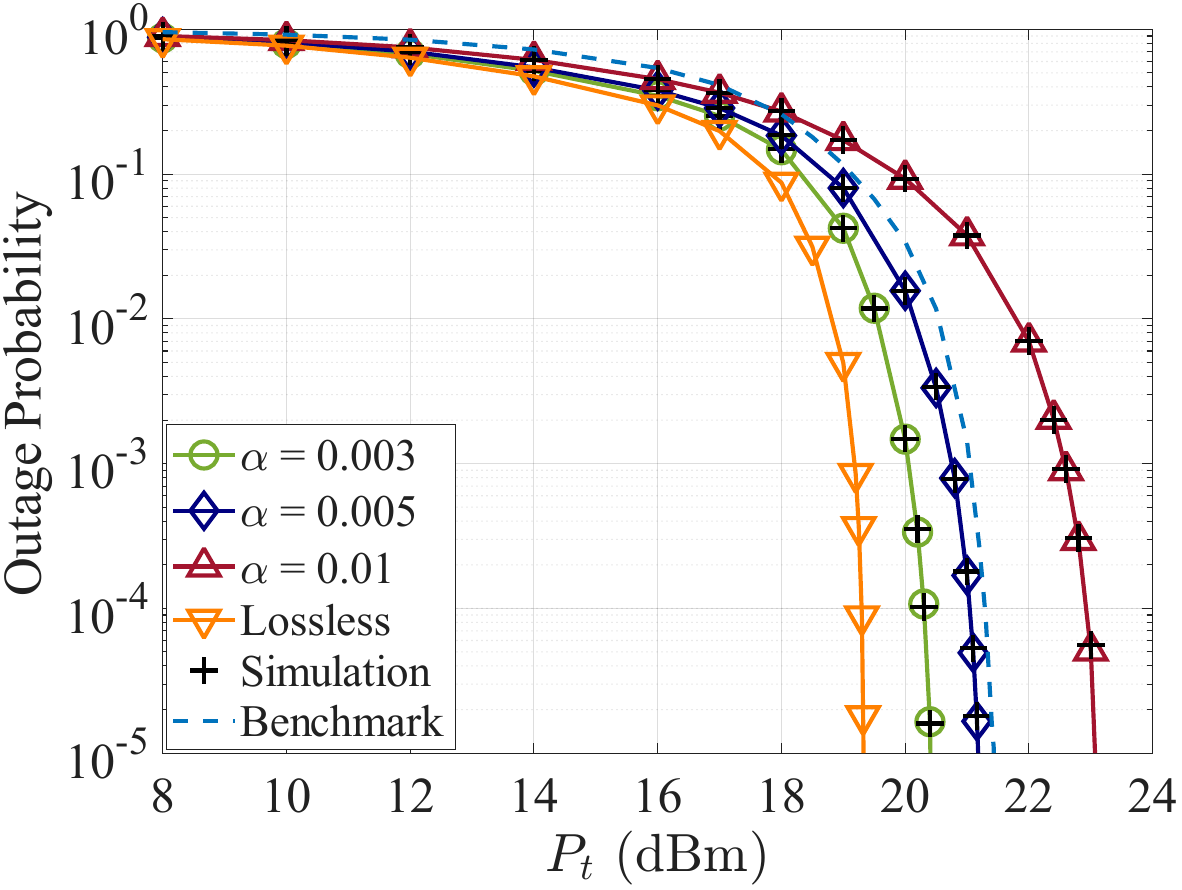}
    \label{fig:OP_Ps_Dxy100}}
\caption{Outage probability versus $P_t$ for different absorption coefficients.}
\label{fig:OP_Ps_Dxy}
\end{figure}

\begin{figure}[t]
\centering
\subfigure[\scriptsize $P_t=12$ dBm.]{
    \includegraphics[width=0.465\columnwidth]{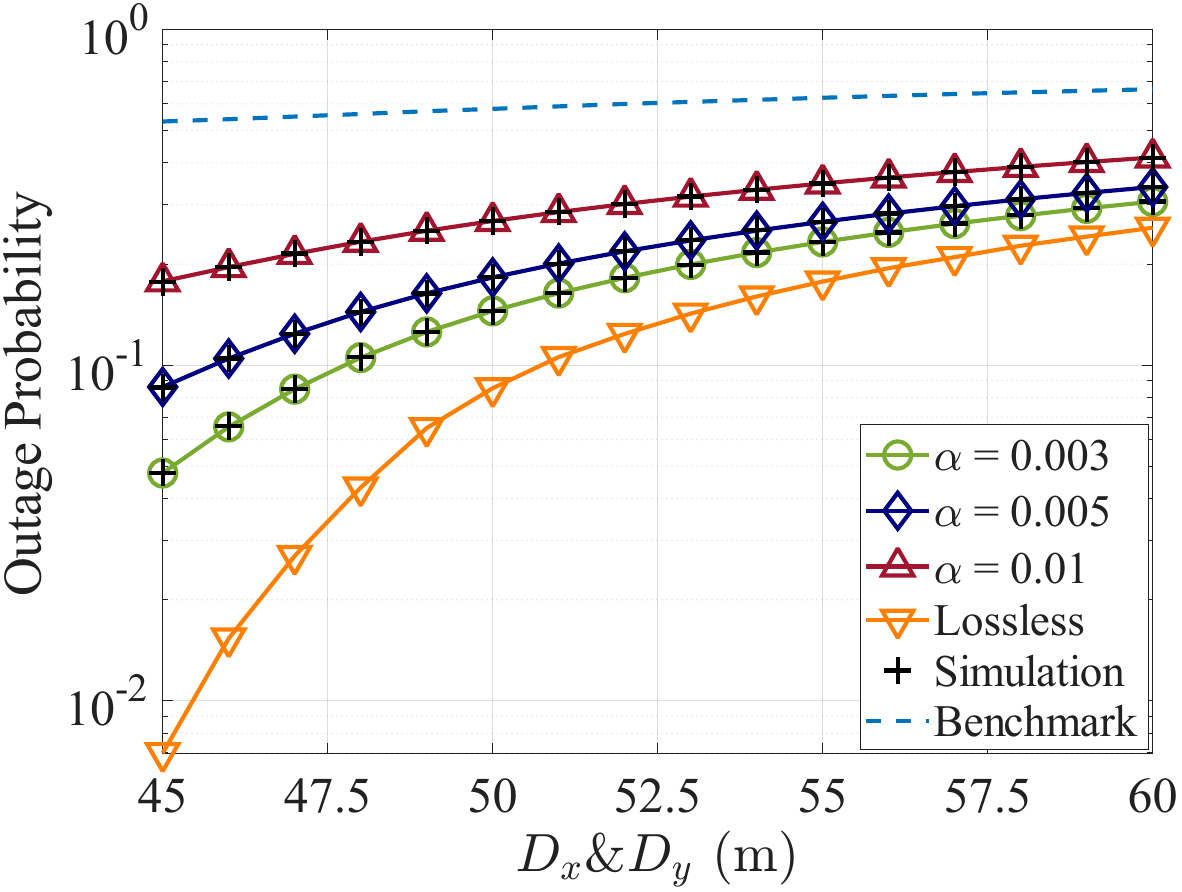}
    \label{fig:OP_Dxy50_Ps12}}
\subfigure[\scriptsize $P_t=18.5$ dBm.]{
    \includegraphics[width=0.465\columnwidth]{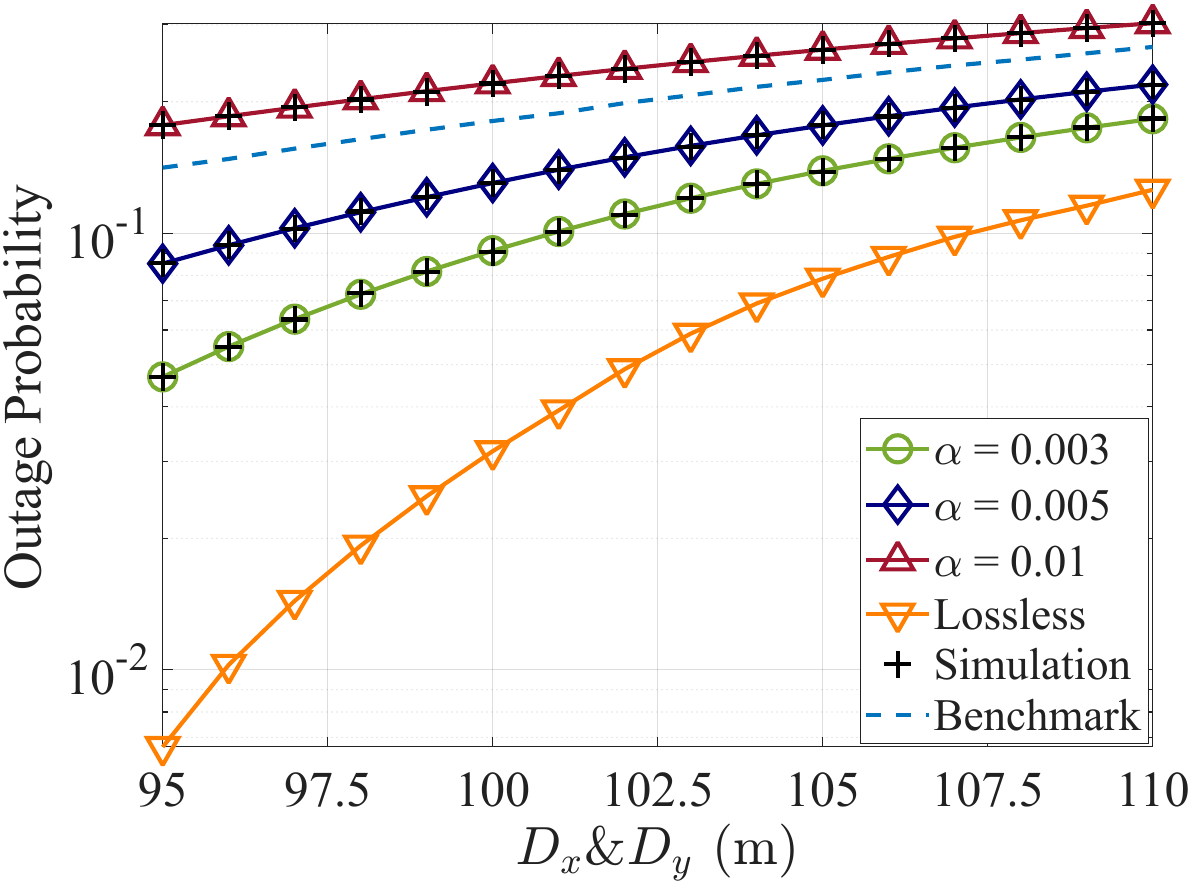}
    \label{fig:OP_Dxy100_Ps18.5}}
\caption{Outage probability versus $D_x$ and $D_y$ for different absorption coefficients.}
\label{fig:OP_Dxy}
\end{figure}



Fig. \ref{fig:OP_Ps_Dxy50} shows the outage probability versus the transmission power $P_t$ for different absorption coefficients, where $D_x=D_y=50$ m. 
For a fair comparison, the H-PAS waveguide is deployed at the same height as the AP in V-PAS.
It is observed that the closed-form expressions of outage probability match well with the corresponding simulation results, verifying the accuracy of our theoretical analysis.
The outage probability of all considered schemes decreases monotonically with increasing $P_t$, and a distinct performance turning point appears around $P_t = 12 \sim 16$ dBm, where the outage probability drops rapidly from the order of $10^{-1}$ down to $10^{-4}$. In addition, the lossless waveguide achieves the lowest outage probability, which constitutes the theoretical performance upper bound of V-PAS. As $\alpha$ increases, the waveguide loss grows exponentially, leading to a significant increase in the outage probability. This performance gap becomes more pronounced in the high $P_t$ regime. Furthermore, the proposed V-PAS exhibits a substantial performance advantage over H-PAS and the benchmark without PA. For the same $P_t$, the outage probability of V-PAS is significantly lower than that of the benchmark, while for a given outage probability requirement, the $P_t$ required by V-PAS is substantially lower. The underlying reason for this advantage is that PA in V‑PAS can be flexibly repositioned along the vertical direction of the dielectric waveguide to always minimize the distance between PA and UAV, effectively reducing the path loss while maintaining a reliable LoS communication link.

Fig. \ref{fig:OP_Ps_Dxy100} shows the outage probability versus the transmission power $P_t$ for different absorption coefficients, where $D_x=D_y=100$ m.
Consistent with the trend observed in Fig. \ref{fig:OP_Ps_Dxy50}, the outage probability of all schemes decreases monotonically with increasing $P_t$. However, due to the expanded UAV operational area, all schemes require higher $P_t$ to achieve the same level of reliability. Among them, the lossless waveguide still achieves the lowest outage probability and constitutes the theoretical performance upper bound of V-PAS. 
It is worth noting that, unlike Fig. \ref{fig:OP_Ps_Dxy50} where all V-PAS schemes outperform the benchmark over the entire $P_t$ range, a clear performance crossing appears in Fig. \ref{fig:OP_Ps_Dxy100}. When $P_t$ is below 18 dBm, the outage probability of all V-PAS schemes is lower than that of the benchmark, maintaining a performance advantage. When $P_t$ exceeds 18 dBm, however, the outage probability of the V-PAS scheme with $\alpha=0.01$ starts to exceed that of the benchmark, and the performance gap gradually widens as $P_t$ further increases. This is because when UAV operational area is large, the PMPL of V-PAS under high $\alpha$ exceeds the direct path loss of the benchmark. This loss difference is more prominently reflected in the outage probability in the high $P_t$ regime.
This phenomenon directly validates the theoretical conclusion in Remark 4. It indicates that the performance advantage of V-PAS is not unconditional and is jointly constrained by multiple key parameters including the absorption coefficient and UAV operational area. In the practical deployment and design of PA, the key parameters should be jointly considered according to specific application scenarios to fully exploit the performance advantages of V-PAS and effectively address the “last meter" coverage problem in communications, rather than ultra-long-distance coverage scenarios.

Fig. \ref{fig:OP_Dxy50_Ps12} and Fig. \ref{fig:OP_Dxy100_Ps18.5} present the outage probability versus the dimensions of UAV operational area $D_x$ and $D_y$ for different absorption coefficients, with $P_t$ fixed at 12 dBm and 18.5 dBm, respectively.
Both figures exhibit a consistent trend. The outage probability of all schemes increases monotonically as $D_x$ and $D_y$ increase. A larger UAV operational area leads to a greater average horizontal distance between PA and UAV, increasing the path loss. The lossless waveguide consistently achieves the lowest outage probability. As $\alpha$ increases, the outage probability increases significantly.
The key distinction between the two figures lies in the relative performance of V-PAS and the benchmark. Fig. \ref{fig:OP_Dxy50_Ps12} corresponds to the condition of a small operational area and moderate $P_t$, where V-PAS exhibits lower outage probability than the benchmark, maintaining a stable performance advantage. Fig. \ref{fig:OP_Dxy100_Ps18.5} corresponds to the condition of a large operational area and high $P_t$, where the V-PAS scheme with $\alpha=0.01$ shows higher outage probability than the benchmark.
These observations verify the theoretical conclusion presented in Remark 4 from the perspective of UAV operational area.

\begin{figure}[t]
\centering
\subfigure[\scriptsize $P_t=12$ dBm, $D_x=D_y=50$ m.]{
    \includegraphics[width=0.465\columnwidth]{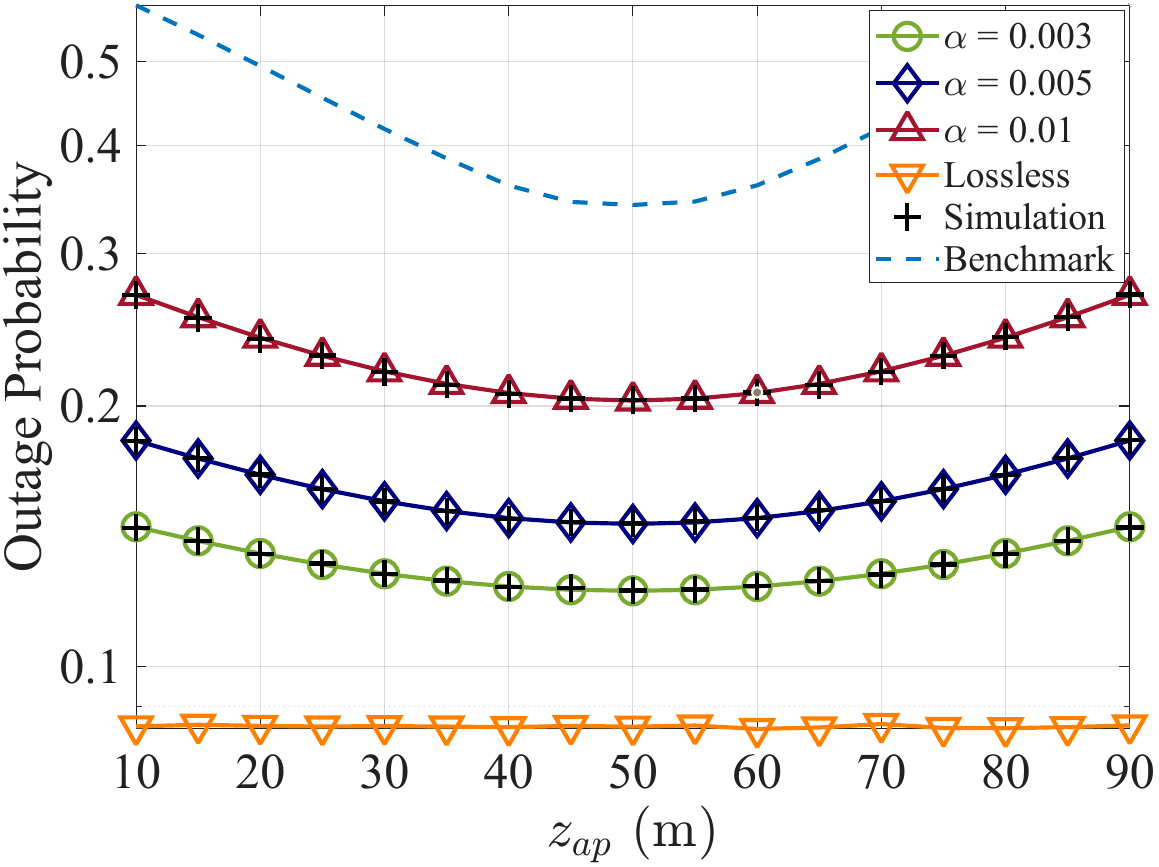}
    \label{fig:OP_hap_Dxy50_Ps12}}
\subfigure[\scriptsize $P_t=16$ dBm, $D_x=D_y=100$ m.]{
    \includegraphics[width=0.465\columnwidth]{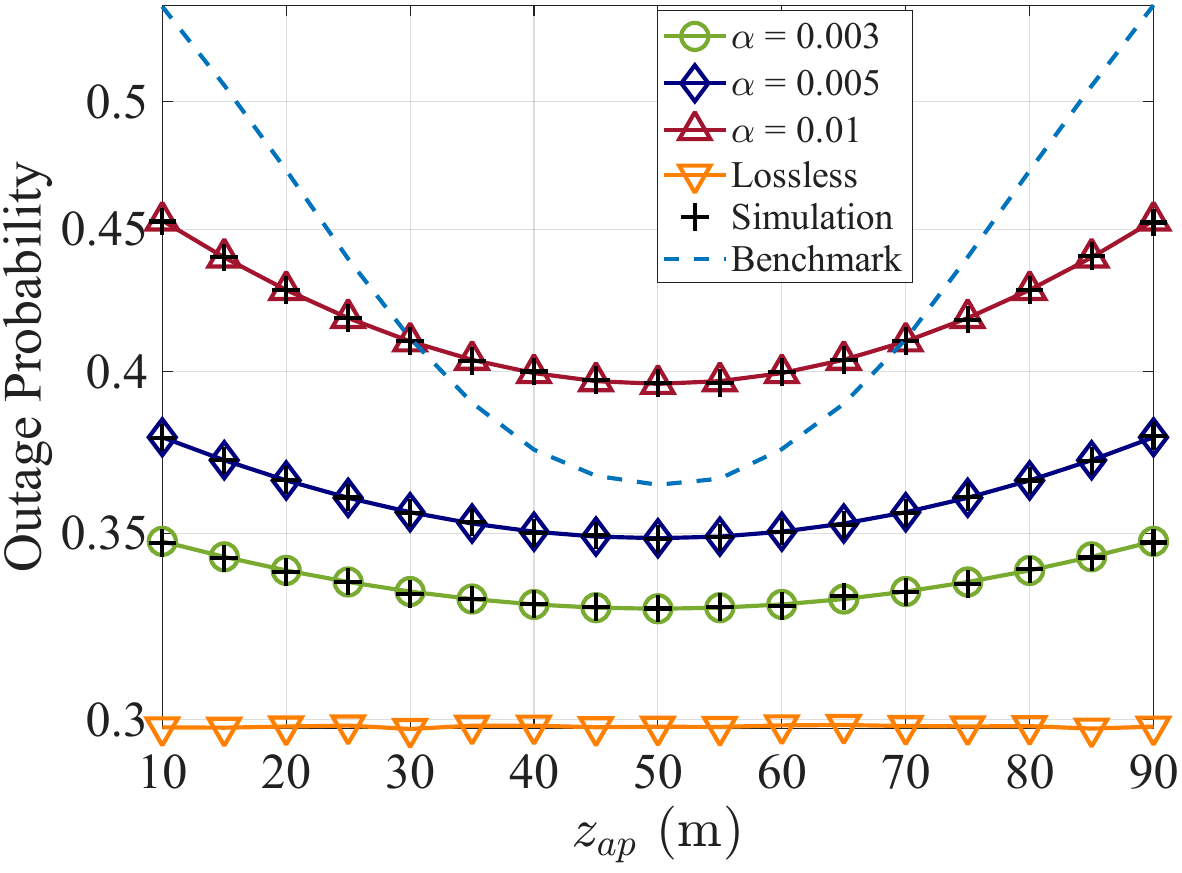}
    \label{fig:OP_hap_Dxy100_Ps16}}
\caption{Outage probability versus $z_{ap}$ for different absorption coefficients.}
\label{fig:OP_hap}
\end{figure}



Fig. \ref{fig:OP_hap_Dxy50_Ps12} and Fig. \ref{fig:OP_hap_Dxy100_Ps16} present the outage probability versus the AP height $z_{ap}$ for different absorption coefficients under $P_t = 12$ dBm, $D_x = D_y = 50$ m and $P_t = 16$ dBm, $D_x = D_y = 100$ m, respectively.
Except for the lossless  waveguide, the outage probability of all schemes is strictly symmetric about $z_{ap}=D_z/2=50$ m, and the minimum outage probability occurs exactly at this midpoint. The outage probability of the lossless waveguide remains the lowest and does not vary with $z_{ap}$. As $\alpha$ increases, the outage probability increases significantly.
The key distinction between the two figures lies in the performance of the benchmark. Unlike Fig. \ref{fig:OP_hap_Dxy50_Ps12}, the outage probability of the benchmark in Fig. \ref{fig:OP_hap_Dxy100_Ps16} is more sensitive to variations in $z_{ap}$. When $z_{ap}$ ranges from $30$ to $70$ m, the benchmark exhibits better reliability than the V-PAS scheme with $\alpha=0.01$.
This symmetry originates from the uniform distribution of UAV vertical position. When $z_{ap}$ is set to $h$ and $D_z-h$, mapping UAV vertical coordinate to $D_z$ minus its original coordinate leaves the statistical distribution of the vertical distance unchanged, resulting in identical outage probabilities for both heights.
The optimal performance at the midpoint stems from the minimization of the average vertical communication distance between AP and PA. When $\alpha = 0$, the waveguide loss is $0$, and the outage probability is independent of the vertical transmission distance. When $\alpha \neq 0$, the average absolute vertical distance is minimized at the midpoint. For the benchmark, the average squared vertical distance is minimized at the midpoint. Therefore, for both the lossy waveguide and the benchmark, the average path loss is minimized at the midpoint, corresponding to the best reliability.
This phenomenon verifies the theoretical conclusion presented in Proposition 3, indicating that there exists a unique optimal AP height that maximizes the reliability. Furthermore, Fig. \ref{fig:OP_hap_Dxy100_Ps16} together with Fig. \ref{fig:OP_Ps_Dxy100} and Fig. \ref{fig:OP_Dxy100_Ps18.5} confirms that the waveguide absorption coefficient, length, and UAV operational area must be jointly considered to maximize V-PAS performance.

\begin{figure}[t]
\centering
\includegraphics[width=0.97\columnwidth]{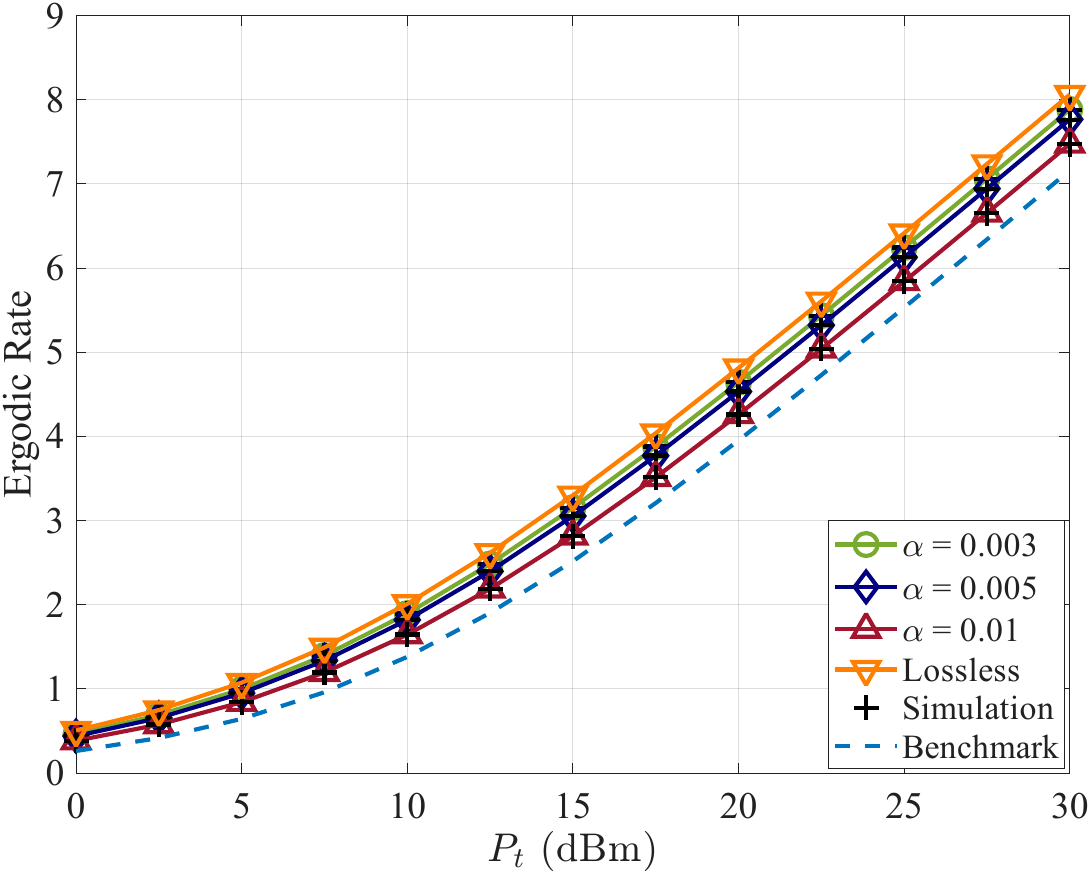}
\caption{Ergodic rate versus $P_t$ for different absorption coefficients, where $D_x=D_y=100$ m.}
\label{fig:AVG_Ps_Dxy100}
\end{figure}

\begin{figure}[t]
\centering
\includegraphics[width=\columnwidth]{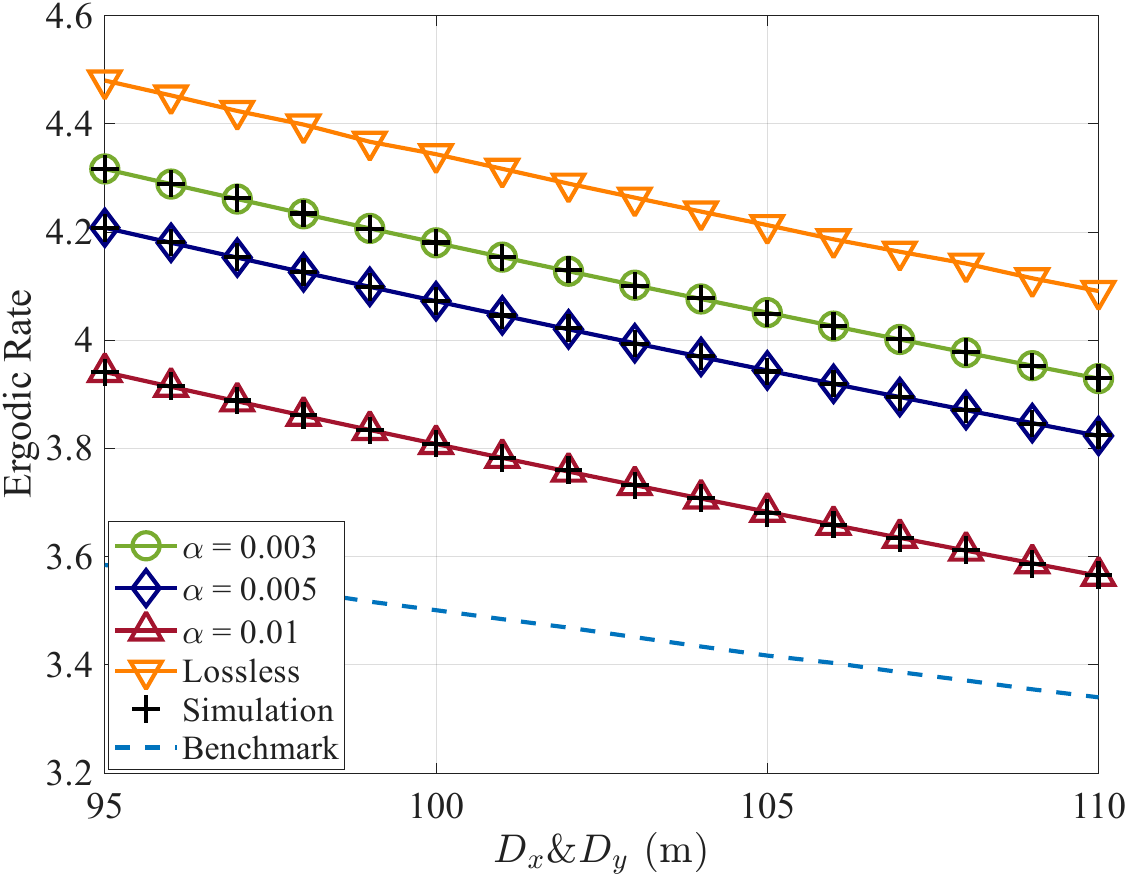}
\caption{Ergodic rate versus $D_x$ and $D_y$ for different absorption coefficients, where $P_t=18.5$ dBm.}
\label{fig:AVG_Dxy100_Ps18.5}
\end{figure}

\begin{figure}[t]
\centering
\includegraphics[width=\columnwidth]{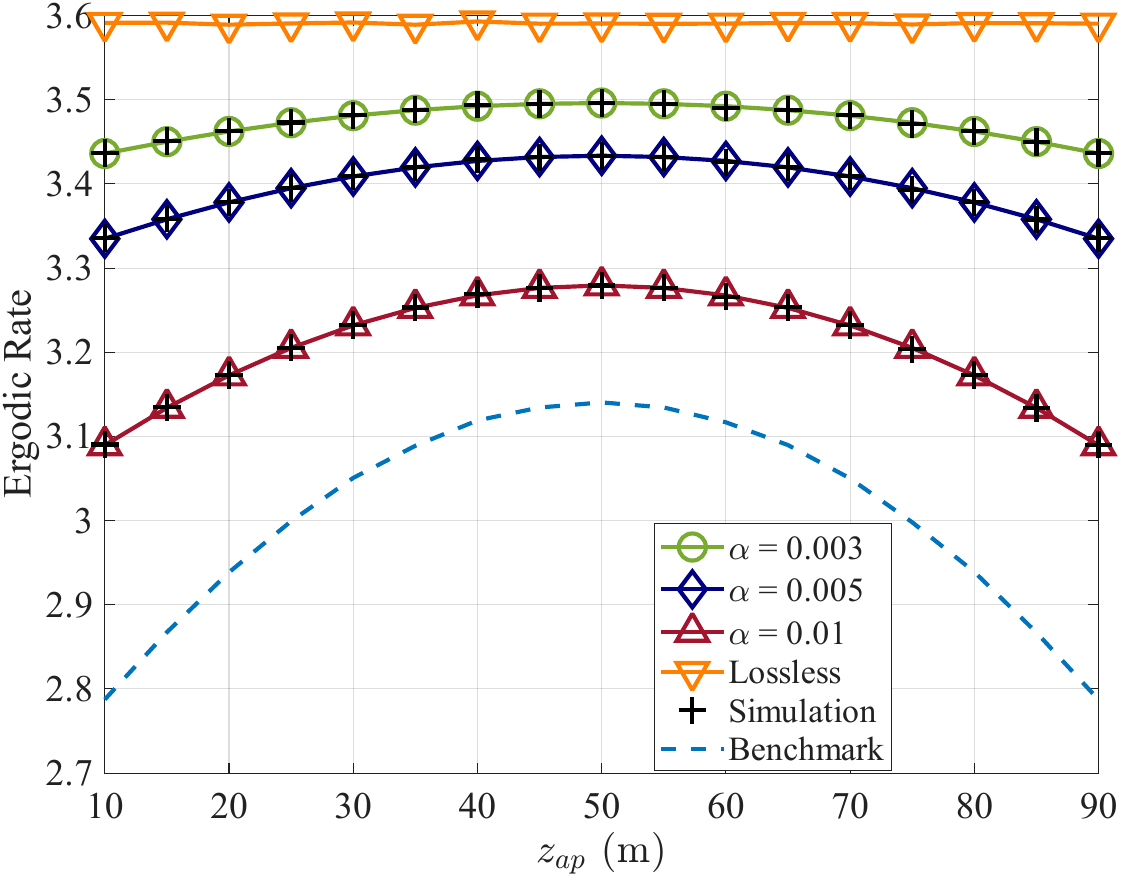}
\caption{Ergodic rate versus $z_{ap}$ for different absorption coefficients, where $P_t=16$ dBm and $D_x=D_y=100$ m.}
\label{fig:AVG_hap_Dxy100_Ps16}
\end{figure}

Fig. \ref{fig:AVG_Ps_Dxy100} shows the ergodic rate versus the transmission power $P_t$ for different absorption coefficients, where $D_x=D_y=100$ m.
It is observed that the closed-form expressions of ergodic rate are in excellent agreement with the corresponding simulation results, verifying the accuracy of our theoretical analysis.
The ergodic rate of all schemes increases monotonically with $P_t$. The lossless waveguide consistently achieves the highest ergodic rate, constituting the theoretical performance upper bound of V-PAS. As $\alpha$ increases, the ergodic rate gradually decreases. However, the ergodic rate of all V-PAS schemes remains higher than that of the benchmark, demonstrating the performance advantage of V-PAS in terms of average throughput.
Interestingly, in Fig. \ref{fig:AVG_Ps_Dxy100}, the V-PAS scheme with $\alpha=0.01$ still outperforms the benchmark in ergodic rate over the entire $P_t$ range, which is in contrast to the outage probability conclusion in Fig. \ref{fig:OP_Ps_Dxy100}. This is because of the different physical natures of the two performance metrics. The outage probability primarily reflects the worst-case performance of the system at low channel capacity, while the ergodic rate, as a statistical average metric, characterizes the average maximum throughput that the system can achieve over the long term.

Fig. \ref{fig:AVG_Dxy100_Ps18.5} shows the ergodic rate versus the dimensions of UAV operational area $D_x$ and $D_y$ for different absorption coefficients, where $P_t=18.5$ dBm.
The ergodic rate of all schemes decreases monotonically as $D_x$ and $D_y$ increase. The lossless waveguide consistently achieves the highest ergodic rate, constituting the theoretical performance upper bound of V-PAS. As $\alpha$ increases, the ergodic rate gradually decreases. However, the ergodic rate of all V-PAS schemes remains higher than that of the benchmark, demonstrating the performance advantage of V-PAS in terms of average throughput.
Even when $D_x=D_y=110$ m, the V-PAS scheme with $\alpha=0.01$ exhibits a higher outage probability than the benchmark, but its ergodic rate still maintains an advantage. This result further confirms the fundamental difference between the two performance metrics.

Fig. \ref{fig:AVG_hap_Dxy100_Ps16} shows the ergodic rate versus the AP height $z_{ap}$ for different absorption coefficients, where $P_t = 16$ dBm and $D_x = D_y = 100$ m.
Similar to the outage probability results in Fig. \ref{fig:OP_hap_Dxy50_Ps12} and Fig. \ref{fig:OP_hap_Dxy100_Ps16}, the ergodic rate of all schemes is strictly symmetric about $z_{ap}=D_z/2=50$ m, and the maximum ergodic rate occurs exactly at this midpoint. This symmetry also originates from the uniform distribution of UAV's vertical position. This result further validates the theoretical conclusion presented in Proposition 3 that there exists a unique optimal AP height that maximizes the ergodic rate performance of V‑PAS.
Furthermore, it is worth noting that the ergodic rate of the benchmark drops more sharply than that of all V‑PAS schemes when $z_{ap}$ deviates from the midpoint. The smaller the absorption coefficient, the more gradual the decrease for  V‑PAS. This indicates that  V‑PAS exhibits greater robustness to variations in $z_{ap}$ compared with the benchmark.
\section{Conclusion}
\label{section:6}
This paper proposed the V-PAS for ultra-low-altitude or low-altitude UAV communications, where the PA was vertically deployed along the facade of buildings to extend the considered user service region of PAS from a 2D ground plane to a 3D low-altitude airspace. Based on the characteristics of the lossy waveguide, the concept of PMPL was defined, revealing its insensitivity to vertical distance and strong adaptability to building heights.
Accurate and asymptotic closed-form expressions for outage probability and ergodic rate under lossy waveguide conditions were derived, and the symmetry and optimality of system performance with respect to the midpoint of AP height were discovered and proved. 
The optimal PA position was further characterized by accounting for the trade-off between waveguide loss and free-space path loss.
The results show that V-PAS outperforms H-PAS and the benchmark without PA in most UAV communications. Only under extreme conditions may the outage probability of V-PAS be inferior to that of the benchmark without PA in UAV communications, while the ergodic rate of V-PAS still maintains an advantage. V-PAS with a lossless waveguide outperforms the benchmark without PA in UAV communications under all conditions, representing the theoretical performance upper bound of V-PAS.

\appendices
\section{Proof of Proposition 1} 
Since $z_m=z_{pa}$, the outage probability can be expressed as
\begin{align}
P_{out}=\mathrm{Pr}(\frac{\eta \rho_te^{-\alpha\left|z_{m}-z_{ap}\right|}}{{x_m}^2+{y_m}^2}<\mathrm{SNR_{thr}}).
\label{eq:OP2}
\end{align}
The outage probability involves integration over all three spatial coordinates.
In the following derivation, we first perform the integration with respect to
$x_m$ and $y_m$, followed by the integration with respect to $z_m$.

Since $x_m$ and $y_m$ are independent and uniformly distributed, their joint probability density function (PDF) is given by
\begin{align}
f(x_m,y_m)=\frac{1}{D_xD_y}.
\label{eq:PDF_xy}
\end{align}
Then, we assume $k=Ce^{-\alpha\left|z_{m}-z_{ap}\right|}$. $P(k)$ is defined as
\begin{align}
P(k)=&\mathrm{Pr}{({x_m}^2+{y_m}^2>k)}\notag\\
=&1-\frac{S(k)}{D_x D_y},
\label{eq:OP3}
\end{align}
where $S(k)$ denotes the area of the intersection between the rectangular region $ [0, D_x] \times [-D_y/2, D_y/2] $ and the circular disk $ x_m^2 + y_m^2 < k $.
In the following, we focus on the practical case 
$D_x > D_y/2$. The case $D_x \leq D_y/2$ can be derived similarly and is omitted for brevity.
Under this assumption, the computation of $S(k)$ can be categorized into four cases. 

\textbf{Case 1:}
For $0 < \sqrt{k} \leq D_y/2$, $S(k)$ of Case 1 is given by 
\begin{align}
S_1(k) = \frac{\pi k}{2}.
\label{eq:S1_k}
\end{align}

\textbf{Case 2:}
For $\frac{D_y}{2}<\sqrt{k}\leq D_x$, $S(k)$ of Case 2 is given by 
\textcolor{blue}{\eqref{eq:S2_k}}.

\textbf{Case 3:}
For $D_x<\sqrt{k}\leq \sqrt{{D_x}^2+\frac{{D_y}^2}{4}}$, $S(k)$ of Case 3 is given by 
\textcolor{blue}{\eqref{eq:S3_k}}.

\textbf{Case 4:}
For $\sqrt{k}> \sqrt{{D_x}^2+\frac{{D_y}^2}{4}}$, $S(k)$ of Case 4 is given by 
\begin{align}
S_4(k) = D_x D_y.
\label{eq:S4_k}
\end{align}

Substituting \textcolor{blue}{\eqref{eq:S1_k}} through \textcolor{blue}{\eqref{eq:S4_k}} into \textcolor{blue}{\eqref{eq:OP3}}, we obtain the expressions for the four cases. These expressions are given by $P_j(k) = 1 - \frac{S_j(k)}{D_x D_y}$, where $j \in \{1,2,3,4\}$.
We further define $K(z_m) = C e^{-\alpha |z_m - z_{ap}|} $. 
Since $z_m$ is uniformly distributed over $[0,D_z]$, 
the outage probability over the 3D service region can be expressed as
\begin{align}
P_{out}=&\mathrm{Pr}\left [ {{x_m}^2+{y_m}^2>K\left ( z_m\right )}\right ]\notag \\
=&\frac{1}{D_z}\int_{0}^{D_z}P\left (K \left (z_m \right )\right )\mathrm{d}z_m.
\label{eq:Pout1}
\end{align}

Next, we need to calculate the outage probability in four cases based on the value of $C$.
To facilitate the calculation, we assume $u\left ( z_m\right )=\left| z_m-z_{ap} \right|$, and $u\left ( z_m\right )\in \left [0,\max(z_{ap},D_z-z_{ap}) \right ]$.

\textbf{Case 1:}
For $C\leq \frac{{D_y}^2}{4}$, $P_{out}$ of Case 1 is given by 
\begin{align}
P_{out,1}=&\frac{1}{D_z}\int_{0}^{D_z}\left [ 1-\frac{\pi K\left ( z_m\right )}{2D_xD_y}\right ]\mathrm{d}z_m\notag \\
=&1-\frac{\pi}{2D_xD_yD_z}\int_{0}^{D_z}Ce^{-\alpha \left|z_m - z_{ap} \right|}\mathrm{d}z_m\notag \\
=&1-\frac{\pi C}{2D_xD_yD_z}\bigg [ \int_{0}^{z_{ap}}e^{-\alpha \left ( z_{ap} - z_m\right )}\mathrm{d}z_m\notag \\
&+\int_{z_{ap}}^{D_z}e^{-\alpha \left ( z_{m} - z_{ap}\right )}\mathrm{d}z_m\bigg ]\notag \\
=&1-\frac{\pi C}{2\alpha D_xD_yD_z}\left [ 2-e^{-\alpha z_{ap}}-e^{-\alpha \left ( D_{z} - z_{ap}\right )}\right ].
\label{eq:P1_out}
\end{align}

\textbf{Case 2:}
For $\frac{{D_y}^2}{4}<C\leq {D_x}^2$, $P_{out}$ of Case 2 is given by 
\begin{align}
P_{out,2}=1-\frac{1}{D_xD_yD_z}\underbrace{\int_{0}^{D_z}S( Ce^{-\alpha u(z_m)})\mathrm{d}z_m}_{I_1}.
\label{eq:P2_out}
\end{align}

Using the substitution $u = \left| z_m - z_{\mathrm{ap}} \right|$, the integral ${I_1}$ with respect to $z_m$ is rewritten as
\begin{align}
I_{1}=\int_{0}^{H_1}S( Ce^{-\alpha u})\mathrm{d}u+\int_{0}^{H_2}S( Ce^{-\alpha u})\mathrm{d}u.
\label{eq:I_1}
\end{align}
Further, we define
\begin{align}
J_i^{(2)}=&\int_{0}^{H_i}S( Ce^{-\alpha u})\mathrm{d}u\notag\\
=&\int_{0}^{m_{1,i}}S_{2}( Ce^{-\alpha u})\mathrm{d}u+\int_{m_{1,i}}^{H_i}S_{1}( Ce^{-\alpha u})\mathrm{d}u
\notag\\
=&\int_{0}^{m_{1,i}}S_{2}( Ce^{-\alpha u})\mathrm{d}u+\frac{\pi C}{2\alpha }\left (e^{-\alpha {m_{1,i}}}-e^{-\alpha {H_{i}}} \right ).
\label{eq:J_2}
\end{align}
Substituting \textcolor{blue}{\eqref{eq:I_1}} and \textcolor{blue}{\eqref{eq:J_2}} into \textcolor{blue}{\eqref{eq:P2_out}}, we have
\begin{align}
P_{out,2}=& 1-\frac{1}{D_xD_yD_z}\displaystyle\sum_{i=1}^{2}\bigg [ \int_{0}^{m_{1,i}}S_{2}(Ce^{-\alpha u})\mathrm{d}u\notag\\
&+\frac{\pi C}{2\alpha }\left (e^{-\alpha {m_{1,i}}}-e^{-\alpha {H_{i}}} \right )\bigg ].
\label{eq:P2_out1}
\end{align}

\textbf{Case 3:}
For ${D_x}^2<C\leq {D_x}^2+\frac{{D_y}^2}{4}$, we define
\begin{align}
J_i^{(3)}=&\int_{0}^{m_{2,i}}S_{3}( Ce^{-\alpha u})\mathrm{d}u+\int_{m_{2,i}}^{m_{1,i}}S_{2}( Ce^{-\alpha u})\mathrm{d}u\notag\\
&+\int_{m_{1,i}}^{H_i}S_{1}( Ce^{-\alpha u})\mathrm{d}u.
\label{eq:J_3}
\end{align}
Then, $P_{out}$ of Case 3 is derived as 
\begin{align}
P_{out,3}=&1-\frac{1}{D_xD_yD_z}\displaystyle\sum_{i=1}^{2}J_i^{(3)}\notag\\
=&1-\frac{1}{D_xD_yD_z}\displaystyle\sum_{i=1}^{2}\bigg [ \int_{0}^{m_{2,i}}S_{3}(Ce^{-\alpha u})\mathrm{d}u\notag\\
&+\int_{m_{2,i}}^{m_{1,i}}S_{2}(Ce^{-\alpha u})\mathrm{d}u+\frac{\pi C}{2\alpha }\left (e^{-\alpha {m_{1,i}}}-e^{-\alpha {H_{i}}} \right )\bigg ].
\label{eq:P3_out}
\end{align}

Subsequently, applying the Gaussian-Chebyshev quadrature, the closed-form expression of $P_{out,3}$ is derived in Table \ref{tab:OP}.

\textbf{Case 4:}
For $C> {D_x}^2+\frac{{D_y}^2}{4}$, we define
\begin{align}
J_i^{(4)}=&\int_{0}^{m_{3,i}}S_{4}( Ce^{-\alpha u})\mathrm{d}u+\int_{m_{3,i}}^{m_{2,i}}S_{3}( Ce^{-\alpha u})\mathrm{d}u\notag\\
&+\int_{m_{2,i}}^{m_{1,i}}S_{2}( Ce^{-\alpha u})\mathrm{d}u+\int_{m_{1,i}}^{H_i}S_{1}( Ce^{-\alpha u})\mathrm{d}u.
\label{eq:J_4}
\end{align}
Then, $P_{out}$ of Case 4 is derived as \textcolor{blue}{\eqref{eq:P4_out}} shown at the top of this page.

Subsequently, applying the Gaussian-Chebyshev quadrature, the closed-form expression of $P_{out,4}$ is derived in Table \ref{tab:OP} and thereby Proposition 1 is proved.
\begin{figure*}[t] 
\begin{align}
P_{out,4}=&1-\frac{1}{D_xD_yD_z}\displaystyle\sum_{i=1}^{2}J_i^{(4)}\notag\\
=&1-\frac{1}{D_xD_yD_z}\displaystyle\sum_{i=1}^{2}\left[D_xD_ym_{3,i}+\int_{m_{3,i}}^{m_{2,i}}S_{3}(Ce^{-\alpha u})\mathrm{d}u
+\int_{m_{2,i}}^{m_{1,i}}S_{2}(Ce^{-\alpha u})\mathrm{d}u+\frac{\pi C}{2\alpha }\left (e^{-\alpha {m_{1,i}}}-e^{-\alpha {H_{i}}} \right )\right ].
\label{eq:P4_out}
\end{align}
\hrulefill
\end{figure*}

\section{Proof of Corollary 1} 
For the asymptotic analysis, as \(P_t \to P_{\mathrm{th}}^-\), only Case 3 should be considered, where \(K \to {T_3}^-\) and \(u \to u_3^+\).

Let \(\varepsilon = T_3 - K\), then \(K = T_3 - \varepsilon\) and \(\varepsilon \to 0^+\). The Taylor expansion of \(S_3(K)\) as \(K \to T_3^-\) is given by
\begin{align}
S_3(K) =& S_3(T_3) + S_3'(T_3)(-\varepsilon) + \tfrac12 S_3''(T_3)(-\varepsilon)^2 + O(\varepsilon^3)\notag\\
=& D_xD_y - \frac{\varepsilon^2}{2D_xD_y} + O(\varepsilon^3),
\label{eq:S3_Taylor} 
\end{align}
where \(O(\cdot)\) denotes the higher-order terms that are omitted, which can be neglected in the asymptotic. 
As $K(u) = C e^{-\alpha u}  = T_3 e^{-\alpha(u-u_3)}$ and \(0 \le u-u_3 \le \max(H_1, H_2)-u_3\),
we have
\begin{align}
e^{-\alpha(u-u_3)} = 1 - \alpha(u-u_3) + O\left((u-u_3)^2\right).
\label{eq:e_Taylor}
\end{align}

Based on \textcolor{blue}{\eqref{eq:e_Taylor}}, \(P(Ce^{-\alpha u})\) can be rewritten as
\begin{align}
P(Ce^{-\alpha u}) &= \frac{(T_3-K)^2}{2D_x^2 D_y^2}\notag \\
&= \frac{T_3^2\alpha^2}{2D_x^2 D_y^2} (u-u_3)^2.
\label{eq:PKasy}
\end{align}

If $z_{ap}\neq \frac{D_z}{2}$, $\int_{u_3}^{\min(H_1, H_2)} P(C e^{-\alpha u})\mathrm{d}u=0$ and the asymptotic expression of outage probability is derived as
\begin{align}
P_{out,asy}&=\frac{1}{D_z} \int_{u_3}^{\max(H_1, H_2)} P(C e^{-\alpha u})\mathrm{d}u \notag\\
&=\frac{T_3^2\alpha^2}{2D_zD_x^2 D_y^2} \int_{u_3}^{\max(H_1, H_2)} (u-u_3)^2\mathrm{d}u \notag\\
&= \frac{T_3^2\alpha^2}{6D_zD_x^2 D_y^2} \left(\max\left(H_1, H_2\right)-u_3\right)^3 \notag\\
&=\frac{\left(D_x^2 + D_y^2/4\right)^2}{6\,\alpha\,D_z\,D_x^2 D_y^2}
\left[\ln\left(\frac{P_{\mathrm{th}}}{P_t}\right)\right]^3.
\label{eq:OP1asy}
\end{align}

If $z_{ap}=\frac{D_z}{2}$, $H_1=H_2$ and the asymptotic expression of outage probability is derived as
\begin{align}
P_{out,asy}=\frac{\left(D_x^2 + D_y^2/4\right)^2}{3\,\alpha\,D_z\,D_x^2 D_y^2}
\left[\ln\left(\frac{P_{\mathrm{th}}}{P_t}\right)\right]^3.
\label{eq:OP2asy}
\end{align}

As \(P_t \to +\infty\), only Case 4 should be considered and $P_{out,asy}=0$.

The proof is completed.

\section{Proof of Proposition 2} 
Based on the definition in \textcolor{blue}{\eqref{eq:Rp}}, the ergodic rate can be expressed as
\begin{align}
R_p = &\frac{1}{D_xD_yD_z}\int_{0}^{D_z}\int_{\frac{-D_y}{2}}^{\frac{D_y}{2}}\int_{0}^{D_x}\notag\\
&\log_{2}{\left ( 1+\frac{\eta \rho_te^{-\alpha\left|z_{m}-z_{ap}\right|}}{{x_m}^2+{y_m}^2}\right )}\mathrm{d}x_m\mathrm{d}y_m\mathrm{d}z_m.\label{eq:Rp3} 
\end{align}
Then, we assume that the horizontal distance between AP and UAV is $r=\sqrt{{x_m}^2+{y_m}^2}$. 
The cumulative distribution function (CDF) of $r$ is given by
\begin{align}
F_R\left ( r\right )=\mathrm{Pr}\left ( R\leq r\right )=\frac{S\left ( r^2\right )}{D_xD_y}.
\label{eq:CDF_r} 
\end{align}
Based on \textcolor{blue}{\eqref{eq:CDF_r}}, the PDF of $r$ is given by
\begin{align}
f_R\left ( r\right )=\frac{1}{D_xD_y}\cdot\frac{\mathrm{d}S\left ( r^2\right )}{\mathrm{d}r}.
\label{eq:PDF_r} 
\end{align}

\textbf{Case 1:}
For $0<r\leq\frac{D_y}{2}$, the PDF of $r$ is expressed as
\begin{align}
f_{R,1}\left ( r\right )=&\frac{1}{D_xD_y}\cdot\frac{\mathrm{d}S_1\left ( r^2\right )}{\mathrm{d}r}\notag\\
=&\frac{\pi r}{D_xD_y}.
\label{eq:PDF_r1}
\end{align}

\textbf{Case 2:}
For $\frac{D_y}{2}<r\leq D_x$, the PDF of $r$ is expressed as
\begin{align}
f_{R,2}\left ( r\right )=&\frac{1}{D_xD_y}\cdot\frac{\mathrm{d}S_2\left ( r^2\right )}{\mathrm{d}r}\notag\\
=&\frac{\pi r-2r\arccos \left ( \frac{D_y}{2r}\right )}{D_xD_y}.
\label{eq:PDF_r2}
\end{align}

\textbf{Case 3:}
For $D_x<r\leq \sqrt{{D_x}^2+\frac{{D_y}^2}{4}}$, the PDF of $r$ can be expressed as
\begin{align}
f_{R,3}\left ( r\right )=&\frac{1}{D_xD_y}\cdot\frac{\mathrm{d}S_3\left ( r^2\right )}{\mathrm{d}r}\notag\\
=&\frac{2r\left ( \arcsin \frac{D_x}{r}-\arccos  \frac{D_y}{2r}\right )}{D_xD_y}.
\label{eq:PDF_r3}
\end{align}

\textbf{Case 4:}
For $r> \sqrt{{D_x}^2+\frac{{D_y}^2}{4}}$, the PDF of $r$ can be expressed as
\begin{align}
f_{R,4}\left ( r\right )=&\frac{1}{D_xD_y}\cdot\frac{\mathrm{d}S_4\left ( r^2\right )}{\mathrm{d}r}\notag\\
=&0.
\label{eq:PDF_r4}
\end{align}

Then, the ergodic rate can be rewritten as
\begin{align}
R_p=\frac{1}{\ln_{}{2}}\int_{0}^{\sqrt{{D_x}^2+\frac{{D_y}^2}{4}}}Q\left ( r\right )f_R\left ( r\right )\mathrm{d}r,
\label{eq:Rp4}
\end{align}
where $Q\left ( r\right )$ is defined as
\begin{align}
Q\left ( r\right ) =& \frac{1}{D_z}\int_{0}^{D_z}\ln_{}{\left ( 1+\frac{\eta \rho_te^{-\alpha\left|z_{m}-z_{ap}\right|}}{r^2}\right )}\mathrm{d}z_m\notag\\
=&\frac{1}{D_z}\Bigg [\underbrace{ \int_{0}^{z_{ap}}\ln_{}{\left ( 1+\frac{\eta \rho_te^{-\alpha\left ( z_{ap}-z_{m}\right )}}{r^2}\right )}\mathrm{d}z_m}_{I_2}\notag\\
&+\underbrace{\int_{z_{ap}}^{D_z}\ln_{}{\left ( 1+\frac{\eta \rho_te^{-\alpha\left ( z_{m}-z_{ap}\right )}}{r^2}\right )}\mathrm{d}z_m}_{I_3}\Bigg ].
\label{eq:Qr}
\end{align}

Using the substitution $u = z_{\mathrm{ap}} - z_m$, the integral ${I_2}$ with respect to $z_m$ can be expressed as
\begin{align}
I_2=\int_{0}^{z_{ap}}\ln_{}{\left ( 1+\frac{\eta \rho_te^{-\alpha u}}{r^2}\right )}\mathrm{d}u.
\label{eq:I_2}
\end{align}

Similarly, using the substitution $u = z_m - z_{\mathrm{ap}}$, the integral ${I_3}$ with respect to $z_m$ can be expressed as
\begin{align}
I_3=\int_{0}^{D_z-z_{ap}}\ln_{}{\left ( 1+\frac{\eta \rho_te^{-\alpha u}}{r^2}\right )}\mathrm{d}u.
\label{eq:I_3}
\end{align}

Observing the forms of $I_2$ and $I_3$, we define a function of $\beta$ as
\begin{align}
G(\beta) = \int_{0}^{\beta} \ln\left(1 + \xi   e^{-\alpha u}\right)\ \mathrm{d}u.
\label{eq:G_beta}
\end{align}
Using the substitution $t = e^{-\alpha u}$ and the properties of the dilogarithm function $\operatorname{Li}_2(x)$, $G(\beta)$ can be expressed as
\begin{align}
G(\beta) =& \frac{1}{\alpha}\int_{e^{-\alpha\beta}}^{1}\frac{\ln(1+\xi  t)}{t}\mathrm{d}t\notag\\
=&\frac{1}{\alpha}\big[-\operatorname{Li}_2(-\xi  t)\big]_{e^{-\alpha\beta}}^{1} \notag\\
=&\frac{1}{\alpha }\big [ \operatorname{Li}_2\left ( -\xi  e^{-\alpha \beta}\right )- \operatorname{Li}_2\left ( -\xi  \right )\big ].
\label{eq:G_beta2}
\end{align}

Therefore, $Q\left ( r\right )$ can be expressed as \textcolor{blue}{\eqref{eq:Q}} and the ergodic rate is derived as \textcolor{blue}{\eqref{eq:Rp5}} shown at the top of this page.
\begin{figure*}[t] 
\begin{align}
R_p=\frac{1}{\ln_{}{2}}\left (\int_{0}^{\frac{D_y}{2}}Q\left ( r\right )f_{R,1}\left ( r\right )\mathrm{d}r+\int_{\frac{D_y}{2}}^{D_x}Q\left ( r\right )f_{R,2}\left ( r\right )\mathrm{d}r+\int_{D_x}^{\sqrt{{D_x}^2+\frac{{D_y}^2}{4}}}Q\left ( r\right )f_{R,3}\left ( r\right )\mathrm{d}r\right ).
\label{eq:Rp5}
\end{align}
\hrulefill
\end{figure*}

Finally, applying the Gaussian-Chebyshev quadrature, the closed-form expression of $R_{p}$ is derived as \textcolor{blue}{\eqref{eq:Rp1}} and Proposition 2 is proved.

\section{Proof of Corollary 2} 
For the asymptotic analysis, as \(P_t \to +\infty\), $1+\frac{\eta P_t e^{-\alpha |z_m-z_{ap}|}}{\sigma^2(x_m^2+y_m^2)}\to\frac{\eta P_t e^{-\alpha |z_m-z_{ap}|}}{\sigma^2(x_m^2+y_m^2)}$ and the asymptotic expression of ergodic rate is given by
\begin{align}
R_{p,asy}=&\frac{1}{\ln2}\mathbb{E}\left[\ln\left(\frac{\eta P_t e^{-\alpha |z_m-z_{ap}|}}{\sigma^2(x_m^2+y_m^2)}\right)\right]\notag\\
=&\frac{1}{\ln2}\Big[\ln P_t+\ln\eta-\ln\sigma^2\notag\\
&-\mathbb{E}\left[\ln(x_m^2+y_m^2)\right]-\alpha\mathbb{E}\left[|z_m-z_{ap}|\right]\Big].
\label{eq:Rp1_asy}
\end{align}
Evaluating these two expectations, we obtain
\begin{align}
\mathbb{E}\left[|z_m-z_{ap}|\right]=&\frac{1}{D_z}\int_0^{D_z}|z_m-z_{ap}|\mathrm{d}z_m\notag\\
=&\frac{z_{ap}^2+(D_z-z_{ap})^2}{2D_z},
\label{eq:Ez_asy}
\end{align}
and
\begin{align}
\mathbb{E}\left[\ln \left(\sqrt{x_m^2+y_m^2}\right)\right]=&\frac{1}{D_xD_y}\int_0^{D_x}\int_{-D_y/2}^{D_y/2}\ln\sqrt{x^2+y^2}\mathrm{d}y\mathrm{d}x\notag\\
=&\frac{1}{D_xD_y}\int_0^{D_x}\bigg[\frac{D_y}{2}\ln\left(x^2+D_y^2/4\right)\notag\\
&-D_y+2x\arctan\left(\frac{D_y}{2x}\right)\bigg]\mathrm{d}x.
\label{eq:Exy_asy}
\end{align}

Based on Eqs. \textcolor{blue}{\eqref{eq:Rp1_asy}}, \textcolor{blue}{\eqref{eq:Ez_asy}}, and \textcolor{blue}{\eqref{eq:Exy_asy}}, we can derive Eqs. \textcolor{blue}{\eqref{eq:Rp_asy}} and \textcolor{blue}{\eqref{eq:Exy}} and Corollary 2 is proved.

\bibliographystyle{IEEEtran}
\bibliography{ref}

@ARTICLE{ZijianZhang,
  author={Zhang, Zijian and Dai, Linglong and Chen, Xibi and Liu, Changhao and Yang, Fan and Schober, Robert and Poor, H. Vincent},
  journal={IEEE Trans. Commun.}, 
  title={Active {RIS} vs. Passive {RIS}: Which Will Prevail in {6G}?}, 
  year={2023},
  volume={71},
  number={3},
  pages={1707-1725},
  month={Mar.},
}

@ARTICLE{Ding,
  author={Ding, Zhiguo and Schober, Robert and Vincent Poor, H.},
  journal={IEEE Trans. Commun.}, 
  title={Flexible-Antenna Systems: A Pinching-Antenna Perspective}, 
  year={2025},
  volume={73},
  number={10},
  pages={9236-9253},
  month={Oct.}}

@ARTICLE{Xu,
  author={Xu, Yanqing and Ding, Zhiguo and Karagiannidis, George K.},
  journal={IEEE Wireless Commun. Lett.}, 
  title={Rate Maximization for Downlink Pinching-Antenna Systems}, 
  year={2025},
  volume={14},
  number={5},
  pages={1431-1435},
  month={May}}

@ARTICLE{WangKaidi,
  author={Wang, Kaidi and Ding, Zhiguo and Schober, Robert},
  journal={IEEE Wireless Commun. Lett.}, 
  title={Antenna Activation for {NOMA} Assisted Pinching-Antenna Systems}, 
  year={2025},
  volume={14},
  number={5},
  pages={1526-1530},
  month={May}}

@ARTICLE{Ouyang,
  author={Ouyang, Chongjun and Wang, Zhaolin and Liu, Yuanwei and Ding, Zhiguo},
  journal={IEEE Trans. Commun.}, 
  title={Rate Region of {ISAC} for Pinching-Antenna Systems}, 
  year={2026},
  volume={74},
  number={},
  pages={5849-5866},
}

@ARTICLE{Wangzhaolin,
  author={Wang, Zhaolin and Ouyang, Chongjun and Mu, Xidong and Liu, Yuanwei and Ding, Zhiguo},
  journal={IEEE Trans. Commun.}, 
  title={Modeling and Beamforming Optimization for Pinching-Antenna Systems}, 
  year={2025},
  volume={73},
  number={12},
  pages={13904-13919},
  month={Dec.}}

@ARTICLE{Ding2,
  author={Ding, Zhiguo and Schober, Robert and Poor, H. Vincent},
  journal={IEEE Trans. Wireless Commun.}, 
  title={Environment Division Multiple Access {(EDMA)}: A Feasibility Study via Pinching Antennas}, 
  year={2026},
  volume={25},
  number={},
  pages={15675-15691},
}

@ARTICLE{Cao1,
  author={Cao, Kunrui and Wang, Tao and Diamantoulakis, Panagiotis D. and Li, Xingwang and Yuen, Chau and Karagiannidis, George K.},
  journal={IEEE J. Sel. Areas Commun.}, 
  title={Reliable and Secure Wireless-Powered Communications via Hybrid Active-Passive Double-{RIS}}, 
  year={2026},
  volume={44},
  number={},
  pages={3828-3844},
}

@ARTICLE{Chen1,
  author={Chen, Jingyu and Cao, Kunrui and Diamantoulakis, Panagiotis D. and Lv, Lu and Yang, Liang and Chi, Haolian and Ding, Haiyang},
  journal={IEEE Trans. Wireless Commun.}, 
  title={Secure Wireless-Powered {zeRIS} Communications}, 
  year={2026},
  volume={25},
  number={},
  pages={225-242},
}

@ARTICLE{Chi,
  author={Chi, Haolian and Cao, Kunrui and Ding, Haiyang and Lv, Lu and Chen, Jingyu and Diao, Danyu and Wang, Buhong and Gong, Fengkui},
  journal={IEEE Internet Things J.}, 
  title={Performance Analysis for {STAR-RIS}-Assisted Wireless Powered Communications With Cooperative Jamming}, 
  year={2025},
  volume={12},
  number={3},
  pages={2574-2591},
  month={Feb.}}

@ARTICLE{Zheng1,
  author={Zheng, Beixiong and Wu, Qingjie and Ma, Tiantian and Zhang, Rui},
  journal={IEEE Trans. Commun.}, 
  title={Rotatable Antenna-Enabled Wireless Communication: Modeling and Optimization}, 
  year={2026},
  volume={74},
  number={},
  pages={6825-6842},
}

@ARTICLE{Zheng2,
  author={Zheng, Beixiong and Ma, Tiantian and You, Changsheng and Tang, Jie and Schober, Robert and Zhang, Rui},
  journal={IEEE Wireless Commun.}, 
  title={Rotatable Antenna Enabled Wireless Communication and Sensing: Opportunities and Challenges}, 
  year={2026},
  volume={33},
  number={3},
  pages={134-141},
  month={Jun.}}

@ARTICLE{D,
  author={Tyrovolas, Dimitrios and Tegos, Sotiris A. and Diamantoulakis, Panagiotis D. and Ioannidis, Sotiris and Liaskos, Christos K. and Karagiannidis, George K.},
  journal={IEEE Trans. Cognit. Commun. Netw.}, 
  title={Performance Analysis of Pinching-Antenna Systems}, 
  year={2026},
  volume={12},
  number={},
  pages={590-601},
}

@ARTICLE{Cao2,
  author={Cao, Kunrui and Chen, Jingyu and Diamantoulakis, Panagiotis D. and Zhou, Lei and Li, Xingwang and Liu, Yuanwei and Karagiannidis, George K.},
  journal={IEEE Trans. Wireless Commun.},
  title={Performance Analysis of Wireless-Powered Pinching Antenna Systems}, 
  year={2026},
  volume={25},
  number={},
  pages={23749-23763},
}

@ARTICLE{Zhangrui1,
  author={Zhu, Lipeng and Ma, Wenyan and Zhang, Rui},
  journal={IEEE Commun. Mag.}, 
  title={Movable Antennas for Wireless Communication: Opportunities and Challenges}, 
  year={2024},
  volume={62},
  number={6},
  pages={114-120},
  month={Jun.}}

@ARTICLE{Zhangrui2,
  author={Liu, Shicong and Yu, Xianghao and Xu, Jie and Zhang, Rui},
  journal={IEEE Trans. Wireless Commun.}, 
  title={Near-Field Communication With Massive Movable Antennas: A Functional Perspective}, 
  year={2026},
  volume={25},
  number={},
  pages={14455-14470},
}

@ARTICLE{KKW1,
  author={Wong, Kai-Kit and Shojaeifard, Arman and Tong, Kin-Fai and Zhang, Yangyang},
  journal={IEEE Trans. Wireless Commun.}, 
  title={Fluid Antenna Systems}, 
  year={2021},
  volume={20},
  number={3},
  pages={1950-1962},
  month={Mar.}}

@ARTICLE{Zhao1,
  author={Zhao, Yizhe and Zhang, Long and Yang, Halvin and Yang, Kun and Zhang, Rui and Song, Lingyang and Liu, Yuanwei},
  journal={IEEE Commun. Surveys Tuts.}, 
  title={Reconfigurable Antennas for Next-Generation Mobile Communication Networks: A Comprehensive Survey and Tutorial}, 
  year={2026},
  volume={28},
  number={},
  pages={5267-5306},
}

@ARTICLE{Shao,
  author={Shao, Xiaodan and Mei, Weidong and You, Changsheng and Wu, Qingqing and Zheng, Beixiong and Wang, Cheng-Xiang and Li, Junling and Zhang, Rui and Schober, Robert and Zhu, Lipeng and Zhuang, Weihua and Shen, Xuemin},
  journal={IEEE Commun. Surveys Tuts.}, 
  title={A Tutorial on Six-Dimensional Movable Antenna for {6G} Networks: Synergizing Positionable and Rotatable Antennas}, 
  year={2026},
  volume={28},
  number={},
  pages={3666-3709},
}

@ARTICLE{Lia,
  author={Liaskos, Christos and Tsioliaridou, Ageliki and Nie, Shuai and Pitsillides, Andreas and Ioannidis, Sotiris and Akyildiz, Ian F.},
  journal={IEEE/ACM Trans. Netw.}, 
  title={On the Network-Layer Modeling and Configuration of Programmable Wireless Environments}, 
  year={2019},
  volume={27},
  number={4},
  pages={1696-1713},
  month={Aug.}}

@ARTICLE{Wu1,
  author={Wu, Zi-Yang and Ismail, Muhammad and Zhang, Jiliang and Zhang, Jie},
  journal={IEEE Wireless Commun.}, 
  title={Tidal-Like Concept Drift in {RIS}-Covered Buildings: When Programmable Wireless Environments Meet Human Behaviors}, 
  year={2025},
  volume={32},
  number={6},
  pages={133-140},
  month={Dec.}}

@ARTICLE{Liu1,
  author={Liu, Fan and Cui, Yuanhao and Masouros, Christos and Xu, Jie and Han, Tony Xiao and Eldar, Yonina C. and Buzzi, Stefano},
  journal={IEEE J. Sel. Areas Commun.}, 
  title={Integrated Sensing and Communications: Toward Dual-Functional Wireless Networks for 6{G} and Beyond}, 
  year={2022},
  volume={40},
  number={6},
  pages={1728-1767},
  month={Jun.}}

@ARTICLE{Vaezi,
  author={Vaezi, Mojtaba and Azari, Amin and Khosravirad, Saeed R. and Shirvanimoghaddam, Mahyar and Azari, M. Mahdi and Chasaki, Danai and Popovski, Petar},
  journal={IEEE Commun. Surveys Tuts.}, 
  title={Cellular, Wide-Area, and Non-Terrestrial {IoT}: A Survey on {5G} Advances and the Road Toward {6G}}, 
  year={2022},
  volume={24},
  number={2},
  pages={1117-1174},
  month={2nd Quart.}}

@ARTICLE{Gran,
  author={Granelli, Fabrizio and Lu, Yan-Ping and Wu, Qingqing and Yuan, Zhenhui and Abdalla, Aly Sabri and Marojevic, Vuk and Jiang, Yifan and Afghah, Fatemeh and Geraci, Giovanni and Mukherjee, Anandarup and Baidya, Sabur and Grieco, Giovanni and Yaqoob, Abid and Hassan, Muhammad Abul and Namuduri, Kamesh and Prasad, Ranga Rao Venkatesha and Lahlou, Laaziz and Sengendo, John and Muntean, Gabriel-Miro},
  journal={IEEE Open J. Commun. Soc.}, 
  title={A Survey on Unmanned Aerial Vehicles ({UAV}s) Communications: State-of-the-Art, Existing Standards, and Future Directions}, 
  year={2026},
  volume={7},
  number={},
  pages={3000-3045},
}

@ARTICLE{Huang,
  author={Huang, Hailong and Su, Jiangcheng and Wang, Fei-Yue},
  journal={IEEE Trans. Intell. Veh.}, 
  title={The Potential of Low-Altitude Airspace: The Future of Urban Air Transportation}, 
  year={2024},
  volume={9},
  number={8},
  pages={5250-5254},
  month={Aug.},}

@ARTICLE{Wangchengxiang,
  author={Wang, Cheng-Xiang and You, Xiaohu and Gao, Xiqi and Zhu, Xiuming and Li, Zixin and Zhang, Chuan and Wang, Haiming and Huang, Yongming and Chen, Yunfei and Haas, Harald and Thompson, John S. and Larsson, Erik G. and Renzo, Marco Di and Tong, Wen and Zhu, Peiying and Shen, Xuemin and Poor, H. Vincent and Hanzo, Lajos},
  journal={IEEE Commun. Surveys Tuts.}, 
  title={On the Road to 6{G}: Visions, Requirements, Key Technologies, and Testbeds}, 
  year={2023},
  volume={25},
  number={2},
  pages={905-974},
  month={2nd Quart.}}

@ARTICLE{Jiang,
  author={Jiang, Yihang and Li, Xiaoyang and Zhu, Guangxu and Li, Hang and Deng, Jing and Han, Kaifeng and Shen, Chao and Shi, Qingjiang and Zhang, Rui},
  journal={IEEE Commun. Mag.}, 
  title={Integrated Sensing and Communication for Low Altitude Economy: Opportunities and Challenges}, 
  year={2025},
  volume={63},
  number={12},
  pages={72-78},
  month={Dec.}}

@article{NTT,
  author={A. Fukuda and H. Yamamoto and H. Okazaki and Y. Suzuki and K. Kawai},
  title={Pinching antenna-using a dielectric waveguide as an antenna},
  journal={NTT DOCOMO Tech. J.},
  volume={23},
  number={3},
  pages={5-12},
  year={2022},
  month={Jan.}}

@book{toolbook,
  author    = {I. S. Gradshteyn and I. M. Ryzhik},
  title     = {Table of Integrals, Series, and Products},
  publisher = {Academic},
  address   = {New York, NY, USA},
  year      = {2007}
}

@ARTICLE{Yangzheng,
  author={Yang, Zheng and Wang, Ning and Sun, Yanshi and Ding, Zhiguo and Schober, Robert and Karagiannidis, George K. and Wong, Vincent W.S. and Dobre, Octavia A.},
  journal={IEEE Wireless Commun.}, 
  title={Pinching Antennas: Principles, Applications and Challenges}, 
  year={2026},
  volume={33},
  number={2},
  pages={175-184},
  month={Apr.}}

@ARTICLE{How,
  author={Tyrovolas, Dimitrios and Tegos, Sotiris A. and Xiao, Yue and Diamantoulakis, Panagiotis D. and Ioannidis, Sotiris and Liaskos, Christos K. and Karagiannidis, George K. and Asimonis, Stylianos D.},
  journal={IEEE Internet Things J.}, 
  title={How Many Pinching Antennas Are Enough?}, 
  year={2026},
  volume={13},
  number={10},
  pages={21994-22006},
  month={May}}

@ARTICLE{GPA,
  author={Xu, Yanqing and Cui, Jingjing and Zhu, Yongxu and Ding, Zhiguo and Chang, Tsung-Hui and Schober, Robert and Wong, Vincent W. S. and Dobre, Octavia A. and Karagiannidis, George K. and Poor, H. Vincent and You, Xiaohu},
  journal={IEEE Commun. Surveys Tuts.}, 
  title={Generalized Pinching-Antenna Systems: A Tutorial on Principles, Design Strategies, and Future Directions}, 
  year={2026},
  volume={28},
  number={},
  pages={5872-5908},
  }

@ARTICLE{5G,
  author={Rappaport, Theodore S. and Sun, Shu and Mayzus, Rimma and Zhao, Hang and Azar, Yaniv and Wang, Kevin and Wong, George N. and Schulz, Jocelyn K. and Samimi, Mathew and Gutierrez, Felix},
  journal={IEEE Access}, 
  title={Millimeter Wave Mobile Communications for {5G} Cellular: It Will Work!}, 
  year={2013},
  volume={1},
  number={},
  pages={335-349},
}

@ARTICLE{ZhaoJJ1,
  author={Zhao, Jingjing and Song, Haowen and Mu, Xidong and Cai, Kaiquan and Zhu, Yanbo and Liu, Yuanwei},
  journal={IEEE Trans. Commun.}, 
  title={Pinching-Antenna Systems-Enabled Multi-User Communications: Transmission Structures and Beamforming Optimization}, 
  year={2026},
  volume={74},
  number={},
  pages={2316-2330},
  month={}
  }

@ARTICLE{ZhaoJJ2,
  author={Zhao, Jingjing and Mu, Xidong and Cai, Kaiquan and Zhu, Yanbo and Liu, Yuanwei},
  journal={IEEE Trans. Wireless Commun.}, 
  title={Waveguide Division Multiple Access for Pinching-Antenna Systems ({PASS})}, 
  year={2026},
  volume={25},
  number={},
  pages={13761-13775},
  }

@ARTICLE{ZhaoJJ3,
  author={Xue, Songtao and Zhao, Jingjing and Cai, Kaiquan and Mu, Xidong and Xiao, Zhenyu and Liu, Yuanwei},
  journal={IEEE Trans. Commun.}, 
  title={Resource Allocation for Pinching-Antenna Systems ({PASS})-Enabled {NOMA} Communications}, 
  year={2026},
  volume={74},
  number={},
  pages={10246-10260},
  }

@ARTICLE{WKD1,
  title={Leaky Coaxial Cable based Generalized Pinching-Antenna Systems with Dual-Port Feeding},
  author={Wang, Kaidi and Ding, Zhiguo and So, Daniel K. C.},
  year={2026},
  note={\emph{arXiv:2602.21856}}
}

@ARTICLE{WKD2,
  title={Leaky-Coaxial Pinching-Antenna System with Adjustable Slot Apertures},
  author={Wang, Kaidi and So, Daniel K. C. and Ding, Zhiguo and Karagiannidis, George K.},
  year={2026},
  note={\emph{arXiv:2604.23246}}
}

@ARTICLE{WKD3,
  title={Pinching-Antenna Systems: From Antenna Placement to Antenna Roaming},
  author={Wang, Kaidi and So, Daniel K. C. and Ding, Zhiguo},
  year={2026},
  note={\emph{arXiv:2608.10136}}
}

@article{ADD1,
  title = {{UAV}-assisted full-duplex {ISAC}: Joint communication scheduling, beamforming, and trajectory optimization},
  journal = {Digit. Commun. Netw.},
  volume = {11},
  number = {5},
  pages = {1628-1638},
  year = {2025},
  author = {Yuanshuo Gang and Yuexia Zhang and Xinyi Wang},
  month={Oct.},
}

@article{ADD2,
  title = {Sum rate maximization in {UAV}-assisted data harvesting network supported by {CF-mMIMO} system exploiting statistical {CSI}},
  journal = {Digit. Commun. Netw.},
  volume = {11},
  number = {4},
  pages = {1280-1292},
  year = {2025},
  issn = {2352-8648},
  author = {Linlin Xu and Qi Zhu and Wenchao Xia and Jun Zhang and Gan Zheng and Hongbo Zhu},
  month={Aug.},
}

@ARTICLE{MC,
  author={Ouyang, Chongjun and Wang, Zhaolin and Liu, Yuanwei and Ding, Zhiguo},
  journal={IEEE Commun. Lett.}, 
  title={Array Gain for Pinching-Antenna Systems ({PASS})}, 
  year={2025},
  volume={29},
  number={6},
  pages={1471-1475},
  month={Jun.},
}

\vfill

\end{document}